\documentstyle{article}
\newcommand{\dir}{.}
\newcommand{\figpage}[2]
{
     \LARGE
     \noindent
     \unitlength=1mm
     \begin{picture}(140,130)
     \put(0,0){
       \psfig{figure=\dir/#2,width=140mm,height=120mm}
     }
     \end{picture}
     \vfill
     \normalsize
     {\tt
       \noindent
       Figure #1 \\
       \figauth \\
     }
   \noindent
}
\begin{document}
\large
\baselineskip=0.8cm
\input psfig
\newcommand{\figauth}{F. Schmid, Physical Review E}
\title{\bf \LARGE Stabilization of Tilt Order by Chain Flexibility 
in Langmuir Monolayers}
\author
{F. Schmid\\
\\
Institut f\"ur Physik, Universit\"at Mainz, D55099 Mainz}
\date{}
\maketitle

{\bf Abstract} - 

Langmuir monolayers are modeled as systems of short chains, which
are confined to a planar surface at one end, but free to move within the plane. 
The phase behavior is calculated in a mean field approximation, which
combines the self consistent field method with elements of classical density 
functional theory. It is shown that phases with tilt order are unstable in 
systems of stiff chains, but can be stabilized by chain conformational entropy 
in systems 
of sufficiently flexible chains. The chain entropy is also responsible
for the appearance of an additional untilted phase, the liquid
expanded phase. The region of stability of the different phases
is discussed, and their microscopic structure is analyzed in some detail.

\vfill

\newpage

\section{Introduction.}

\bigskip
Monolayers of amphiphilic molecules have been studied for many years
for practical and fundamental reasons. Placed on a solid substrate, 
they build Langmuir-Blodgett films, which have important technical 
applications, {\em e.g.}, in thin film technology \cite{roberts}. 
Monolayers of lipids on water are of 
biological interest, since lipid bilayers -- consisting of two weakly
coupled monolayers -- are essential ingredients of biological
membranes \cite{gennis}. 

The phase diagram of Langmuir monolayers (monolayers adsorbed
at the air water interface) at low surface coverage 
is qualitatively similar for long chain fatty 
acids, alcohols and lipids (Figure 1) \cite{knobler90,mono-g}. It's most
remarkable feature is the presence of two distinct fluid-fluid
coexistence regions at intermediate temperatures: the familiar
``gas -- liquid'' transition at low surface densities, and an additional
transition from a ``liquid expanded'' (LE) phase to a ``liquid condensed'' 
state at higher densities. The latter is indeed a transition between two 
fluid states, as evidenced by the experimental observation that positional 
correlations in the condensed phase decay exponentially \cite{helm87}.
It is the monolayer equivalent of the ``main'' transition in bilayers, which
is interesting from a biological point of view, because it is found
at temperatures often close to the body temperature (41.5 ${}^0$C in DPPC)
\cite{bi}. At even higher surface coverage, monolayers can
display a rich spectrum of condensed phases, which differ from each other in
positional order, tilt order, and orientational order of the backbones of the 
chains \cite{mono-p}. In this work, we shall discuss the condensed phases 
which can coexist with the expanded phase, {\em i.e.} the high temperature
untilted phase (LS) and the low temperature tilted phase 
(L${}_2$, see Figure 1).

The nature of the transition between liquid expanded and condensed phases
has been discussed over many years. In an earlier paper \cite{me1}, we
have presented self consistent field calculations of a ``minimal'' model 
for Langmuir monolayers, where the amphiphilic molecules were modeled as 
semiflexible chains with one end grafted to a planar surface. We have
shown that two ingredients are needed to bring about coexistence between
two liquid states: The chain flexibility, which stabilizes the expanded
phase, and the chain anisotropy, which dominates the liquid condensed state.
The transition is driven by the interplay between the entropy of chain
disorder and the energy associated with collective chain alignment.
The latter may result from simple packing effects, or from additional
({\em e.g.}, dipolar) anisotropic interactions between chain segments.

The model hence successfully reproduced the LE and the LS phase, yet
it seemed to fail to display stable phases with collectively tilted chains. 
Indications for tilt order were only seen in the unstable regions
of two phase coexistence. In that respect, the observed phase behavior
was similar to that of grafted rigid rod systems. Grafted rods with fixed
grafting points may show tilt order in a region of surface coverage
\cite{wang,scheringer}. However, the surface energy per chain is higher in 
the tilted region than in the untilted region. When the rods are given
translational degrees of freedom, the tilting transition is therefore
replaced by phase separation \cite{shin,li}. 

According to a common picture, tilt order in Langmuir monolayers 
results from a mismatch between head group and tail segment size. The 
larger area of the head group constrains the coverage of the condensed phase 
and stabilizes surface coverage regions with tilt order. This mechanism is
doubtless the driving force for tilt order in many cases, but it is
certainly not the only one. For example, it hardly explains the
experimental observation of tilt order in monolayers of triple chain 
phospholipids \cite{dietrich}. 
Another potential cause for tilt is related to the internal structure of the
chains: When the chains are tilted, monomers can ``hook'' into each other, 
and thus pack more effectively. Presumably, this is responsible 
for the presence of tilt order in Monte Carlo simulations of endgrafted
bead-spring-chains \cite{frank, harald}. 
Tilt order may also be induced by attractive interactions between the chains 
and the bare surface \cite{wang,Ahalperin}. 
For hydrophobic ((CH${}_2$)${}_n$) chains on a water
surface, that seems however less likely.

All these tilting mechanisms do not operate in the minimal model of
reference \cite{me1}. Therefore one would not expect to find tilt order there,
unless the model is extended in a suitable way. Yet we shall show that 
the conformational degrees of freedom of the chains generate a new
mechanism for the stabilization of tilted states: As we have discussed
above, giving the chains some flexibility brings a new phase into
existence, the liquid expanded phase. On making the chains more and more
flexible, the condensed phase is affected too: The gain of conformational
entropy at lower surface densities compensates in part the loss of surface
energy. As a result, the region of stability of the condensed phase is
extended. Provided the chains are sufficiently anisotropic, the coverage
at coexistence becomes low enough to support collective tilt. 

The influence of conformational chain disorder on tilt in fatty acid 
monolayers has received some interest recently \cite{karaborni2,li2}. 
It has been argued that in tilted phases, an increase in the number of 
gauche defects in the chains reduces the tilt
angle at the same area coverage. The present work discusses an antipodal,
although related effect: Chain disorder stabilizes homogeneous tilted 
phases at molecular areas, where ordered, straight, chains phase separate into
two untilted phases.

The purpose of this work is two fold: To explore the possibilities
for tilt order within the minimal model, and to establish a 
complete phase diagram in terms of the variables stiffness and chain 
anisotropy. The parameter region, in which tilted phases are stable,
will be determined, as well as the parameter region, in which a liquid
expanded and a liquid condensed phase can coexist. Where those two
regions overlap, one finds a phase diagram which is very similar to the 
one sketched in Figure 1. 
The paper is organized as follows. The model and the self consistent field
method are described in the next section. A variant of the model \cite{me1}
is used, which allows among other for a more detailed study of chain
defects. Section three presents the predictions of the self consistent field
theory first for flexible chains, then for stiff chains. 
The properties of the different phases are discussed in some detail 
(density profiles, nematic order, chain defects), and an overview over 
the phase behavior is given. The results are summarized in section four.

\bigskip

\section{ The Model.}

\bigskip

A schematic picture of the model is shown in Figure 2. The amphiphilic
molecules are modeled as chains containing $n$ rod-like tail segments 
of length $l_0$ and diameter $A_0$, and one head segment, which is
confined to a planar surface at $z=0$. 
They are subject to three different types of potentials: 
\begin{itemize}
\item External potentials, which confine the head segment at $z<0$ and
the tail segments at $z>0$
\item A bending potential, which favors parallel alignment of
adjacent segments.
\item The interactions between segments. 
Tail segments are anisotropic, have a repulsive hard core,
and attract each other at larger distances.
Head segment interactions are isotropic and purely repulsive.
\end{itemize}

The external potentials $h_h^{ext}$ (head segments) and $h_t^{ext}$ 
(tail segments) are taken to be simply harmonic.
\begin{equation}
\frac{h_h^{ext}(\vec{r})}{k_B T}
 = \left\{ \begin{array}{ll} 0 & z<0\\k_h z^2 & z>0 \end{array}
\right.
\qquad
\mbox{and}
\qquad
\frac{h_t^{ext}(\vec{r})}{k_B T}
 = \left\{ \begin{array}{ll} k_t z^2 & z<0\\0 & z>0 \end{array}
\right.,
\end{equation}
where $k_B$ is the Boltzmann constant and $T$ the temperature.

The choice of the bending potential is guided by the idea that 
tail segments in the model chain correspond to two CH${}_2$ groups each in a 
hydrocarbon chain. A molecule in an {\em all trans} conformation is
then represented by a completely stretched model chain with 
bending angles $theta = 0$.
A conformation with one {\em gauche} kink is represented by a model chain,
which has one bending angle $\theta = (\pi/3)$ or $\cos \theta = (1/2)$.
In view of these considerations, the bending potential is given the
the form $U(\theta)/(k_B T) = u \widehat{U}(\theta)$, where $u$ is an
adjustable stiffness parameter, and
\begin{equation}
\label{u}
\widehat{U}(\theta) = 25 x + 34 x^2 - 400 x^3 + 480 x^4 
\qquad \mbox{with} \qquad
x = 1 - \cos \theta.
\end{equation}
The function $\widehat{U}(\theta)$ is plotted in Figure 2. It has a minimum
at $\cos(\theta)=1/2$, and takes the value $U_0=1$ there. The 
relative potential barrier $U_m/U_0 \approx 4 $ has approximately the same 
height as
the energy barrier from {\em trans} to {\em gauche} in popular polyethylene 
models (e.g. by Rigby and Roe \cite{rigby}). Moreover, the thermal 
average of $\cos(\theta)$ in a free model chain at $u = 1$, 
$\langle \cos (\theta) \rangle \approx 0.7$, is in rough agreement with the 
value obtained for polyethylene at $k_B T = E_g$, where $E_g$ is the
energy of a {\em gauche} defect (calculated in the RIS scheme, \cite{RIS}).

The mapping of carbon groups on chain segments should not be taken
too literally, since the model is so simple compared to a real hydrocarbon
chain. However, the choice of a bending potential with two minima such 
as (\ref{u}) has the advantage, that it allows to define chain defects, and 
to study defect distributions. Note that the energy of {\em gauche } 
defects is of order $300 K$, {\em i.e.}, room temperature, 
in units of the Boltzmann constant $k_B$. Hence an analysis of their
distribution can be instructive, especially in short chains.
For most other purposes, a simple harmonic potential such as has been used 
in Ref. \cite{me1} is entirely sufficient, and yields qualitatively the same 
results.

The interaction between segments are introduced in terms of a functional 
${\cal F}
[\{ \widehat{\rho}_h(\vec{r},\vec{w}),\widehat{\rho}_t(\vec{r},\vec{w}) \}]$ 
of the center of mass densities of head ($\widehat{\rho}_h$) and tail 
($\widehat{\rho}_t$) segments with orientation $\vec{w} \: (|\vec{w}|=1)$
at position $\vec{r}$. 
The functional includes short range repulsive hard core potentials
as well as longer range attractive interaction tails. 

In an exact treatment of the above model, one has to perform ensemble 
averages over all possible
configurations of chains, and the corresponding center of mass densities. 
In this work, we will resort to a local mean field approximation. The densities
$\widehat{\rho}_{h,t}(\vec{r},\vec{w})$ are replaced by their ensemble 
averages, and ${\cal F}$ is taken to be a functional of average 
densities. Single segments interact with others {\em via} average fields
\begin{equation}
\label{meanfield}
h^{ind}_{h,t}(\vec{r},\vec{w}) = 
\frac{\delta {\cal F}}
{\delta \widehat{\rho}_{h,t}(\vec{r},\vec{w}) }.
\end{equation}
The effect of local density fluctuations is neglected.

This approximation has a number of important implications.
First, correlations between different chains are neglected.
The problem therefore reduces to calculating the partition function and the
density distribution of a single noninteracting chain (random walk) in the 
inhomogeneous external fields 
$h_{h,t}(\vec{r},\vec{w}) = h_{h,t}^{ind} + h_{h,t}^{ext}$,
which have to be determined self consistently using eqn (\ref{meanfield})
(cf. \cite{scheutjens,szleifer}). 
Note that correlations within a chain are still present 
due to the chain connectivity.
Second, the non integrable hard core interactions require special treatment.
We will choose a common approach in density functional theories
\cite{coy2,me3},
which is to expand around a reference system of purely repulsive segments. 
Third, the mean field approximation does not capture the fact that
stiff chains are always anisotropic, even if the constituting segments are not. 
Effective anisotropic interactions result, {\em e.g.}, from packing 
effects. Within the mean field approach, they have to be introduced explicitly
in terms of an effective segment anisotropy.

Since the segments are extended objects, the center of 
mass density $\widehat{\rho}$ corresponds to a segment mass density
\begin{equation}
\rho_{h,t}(\vec{r},\vec{w}) =
\int d \vec{r}\,' K_{h,t}(\vec{r}-\vec{r}\,',\vec{w}) 
\widehat{\rho}_{h,t} (\vec{r}\,',\vec{w}).
\end{equation}
The function $K(\vec{r},\vec{w})$ reflects the shape of a segment with
orientation $\vec{w}$. 
Since the segments are fairly compact, the orientation dependence of
the shape function $K(\vec{r},\vec{w})$ can be neglected, and it is
reasonably well approximated by a simple step function.
\begin{equation}
K(\vec{r},\vec{w}) = 
\left\{
\begin{array} {rl} 1/(l_0 A_0) & |z|< l_0/2, \; (x^2+y^2) < A_0/\pi \\
0 & \mbox{otherwise}
\end{array}
\right.
\end{equation}

We shall also need the total density
\begin{equation}
\rho(\vec{r}) = \frac{1}{4 \pi} \int d\vec{w} \:
[ \rho_t(\vec{r},\vec{w}) + \rho_h(\vec{r},\vec{w})],
\end{equation}
where the integral $ \int d\vec{w} $ is performed over the full solid 
angle $4 \pi$. With these definitions, we are able to formulate a concrete
Ansatz for the density functional {\cal F}. We use a local density
approximation, {\em i.e.}, the functional {\cal F} is given as the
integral over a free energy density function. 
\begin{equation}
\label{ff}
\frac{1}{k_B T} {\cal F} =
\int d \vec{r} \Big\{
f_0[\rho(\vec{r})] +
\frac{1}{32 \pi^2} \int \int d\vec{w} \; d\vec{w}'
\rho_t(\vec{r},\vec{w}) \rho_t(\vec{r},\vec{w}')
[V(\vec{w} \cdot \vec{w}') - e]
\Big\}
\end{equation}

The first term describes a reference system of identical segments 
with isotropic hard core interactions. The free energy density 
$f_0[\rho]$ is derived from the hypothetical equation of state of a 
dense melt of such ``ideal'' chain segments:
Being part of a chain, the segments have no translational degrees of freedom,
their equation of state has no ideal gas contribution. Furthermore, segments
are connected to others at both ends, therefore they mainly interact 
within a plane perpendicular to themselves.
Hence we assume that their equation of state is reasonably well
approximated by the equation of state for hard disks \cite{disks}, 
from which the ideal gas term has been subtracted.
\begin{equation}
\Pi(\rho) = \rho (\frac{1}{(1 - \eta)^2} - 1)
\end{equation}
with the reduced pressure $\Pi=p/k_B T$, the density $\rho$ and the
packing fraction $\eta = \rho \cdot A_0 l_0$. From this one
can derive the free energy density using $d(f_0/\rho)/d\rho = \Pi/\rho^2$.
\begin{equation}
f_0[\rho] = \rho \{\frac{\eta}{1-\eta} - \log(1-\eta)\}
\end{equation}

The second term in eqn (\ref{ff}) accounts for the anisotropic and attractive 
interactions between tail segments perturbatively, up to the leading 
order in the densities. The attractive part of the interaction is 
absorbed in a single parameter $e$. The anisotropic part of the interaction is
described by an even function $V(x)=V(-x)$ and can be expanded in
Legendre polynomials.
\begin{equation}
V(x) = V(-x) = \sum_{l=2,4,\cdots}^{\infty} \frac{2l+1}{4 \pi} P_l(x) v_l.
\end{equation}
We shall neglect all contributions except for the lowest, $v \equiv - v_2$.
It should be emphasized again that the anisotropy parameter, $v$, cannot
necessarily be traced back to an actual anisotropy of single free segments.
It is an effective parameter, which has to be introduced in a mean field
theory in order to include effects of the chain anisotropy. Thus it has to be
identified with an effective anisotropy {\em per} segment, rather than with 
the anisotropy {\em of} a segment.

We complete the definition of the model by specifying the parameters
\mbox{
$k_h=k_t=20/3\, l_0^{-2}$}, \mbox{$e=40\, l_0^3$}, \mbox{$A_0=2.01\, l_0^2$} 
(see \cite{me1}).
This choice is motivated as follows: 
The parameters $k_h$ and $k_t$ can be chosen arbitrarily, provided they are
large enough to ensure the confinement of the heads at the surface,
and of the chains above the surface. The strength of the attractive 
interaction, $e$, determines the density within a hydrophobic
layer, and affects the jump in the surface coverage at first order
transitions. As shown in reference \cite{me1}, $e$ has not much 
qualitative influence on the phase behavior, therefore it is not
varied systematically here. In a virial expansion, $e$ is given by
the integral over the Mayer-f-function of the attractive interaction,
\mbox{$e = \int (\exp[- v_{attr.}(\vec{r})/k_B T]-1) d\vec{r}$}. The parameters
$e$ and the effective chain diameter $A_0$ were chosen such that they
are compatible with the size and potentials of alkane chains, if one
maps two $(CH_2)_n$ groups on one model segment, with alkane potentials
taken from ref. \cite{coy3}. 

All calculations were done with chains of tail length $n=7$.
Free model parameters, which were systematically varied, are
the stiffness parameter $u$ and the anisotropy parameter $v$,
hereafter given in units of $l_0^3$.
We shall comment briefly on their connection with interaction
parameters in other systems, {\em e.g.}, simulation models.
In a simulation, the effective stiffness $u$ can be estimated from 
matching the thermal average $\langle \cos \theta \rangle$ 
for the angle $\theta$ between adjacent bonds, 
in a dense melt of free chains, 
with the average obtained for a random walk of rods with
the bending potential $u \widehat{U}$ (\ref{u}). The least accessible
parameter is the effective anisotropy parameter $v$. As noted earlier,
it's origin is mostly due to packing effects. A lower bound can be
calculated from the excluded covolume (the second virial coefficient) of 
two stretched chains of the persistence length, 
divided by the number of segments. 
In the present model, at $u \sim 1-2$, one gets $v \sim 10 l_0^3$. 
Such a calculation however neglects the anisotropy in the attractive
interaction. Moreover, the segment density in the hydrophobic layer is
very high (see Figure 10), such that higher order virial coefficients
come heavily into play. 
Hence the resulting effective anisotropy will be much higher. 
In a simulation, $v$ can be determined from the analysis of 
orientation correlations between segments in a melt of free chains.

The procedure used to solve the problem is similar to the Scheutjens-Fleer
method for lattice models of polymers at surfaces \cite{scheutjens}. 
One defines recursively the end segment distributions ($i \le n$)
\begin{eqnarray}
W_i(\vec{r},\vec{w}) &=& \frac{1}{4 \pi} \int d \vec{w}'
W_{i-1}(\vec{r}\,',\vec{w}')
e^{(- h_t(\vec{r},\vec{w})-U( \vec{w}\cdot \vec{w}'))/k_B T}  \nonumber\\
&& \qquad
\vec{r}\,' = \vec{r} - \frac{l_0}{2}[\vec{w} + \vec{w}'] 
\\
\overline{W}_i(\vec{r},\vec{w}) &=& \frac{1}{4 \pi} \int d \vec{w}'
W_{i+1}(\vec{r}\,',\vec{w}')
e^{(- h_t(\vec{r}\,',\vec{w})-U( \vec{w}\cdot \vec{w}'))/k_B T}  \nonumber\\
&&\qquad 
\vec{r}\,' = \vec{r} + \frac{l_0}{2}[\vec{w} + \vec{w}'] 
\end{eqnarray}
with $W_0(\vec{r},\vec{w})=\exp(-h_h(\vec{r})/k_B T)$ and
$\overline{W}_n(\vec{r},\vec{w})=1$. We consider a homogeneous monolayer of 
$N$ chains, which occupy each an area per molecule $A$. Hence we have
translational invariance on the $xy$ plane, and the 
single chain partition function is given by
\begin{equation}
{\cal Z}_0 = \frac{1}{4 \pi l_0} \int dz \: d\vec{w} \:
W_i(z,\vec{w}) \overline{W}_i(z,\vec{w}),
\end{equation}
which is independent of $i$.
The center of mass density of the $i$th segment can be calculated {\em via}
\begin{equation}
\widehat{\rho}_i(z,\vec{w})
= \frac{1}{A l_0} \; 
\frac{W_i(z,\vec{w})\: \overline{W}_i(z,\vec{w})}{{\cal Z}_0} 
\end{equation}
and the total free energy per chain 
(with the de Broglie wavelength $\lambda_B$)
\begin{equation}
\label{ffree}
\frac{F}{N k_B T} = - \log {\cal Z}_0 - \log(A/\lambda_B^2) - 1 
\end{equation}
\begin{displaymath}
+ A \int dz \; \Big\{ 
(f_0[\rho] - \rho \frac{d f_0}{d \rho}) - \frac{1}{32 \pi^2}
\int d\vec{w}\: d\vec{w}' \rho_t(z,\vec{w}) \rho_t(z,\vec{w}')
(V(\vec{w} \vec{w}') - e) \Big\}.
\end{displaymath}
The chemical potential $\mu$, 
{\em i.e.}, the free energy gain on adding one chain is
therefore given by
\begin{equation}
\frac{1}{k_B T} \; \frac{\partial F}{\partial N} = 
\frac{\mu}{k_B T} = - \log \frac{{\cal Z}_0 A}{\lambda_B^2}.
\end{equation}
In the grand canonical ensemble, this leads to the Gibbs free energy
per surface area $g$
\begin{equation}
\label{gg}
\frac{g}{k_B T} = \frac{1}{A}(\frac{F}{N k_B T} - \frac{\mu}{k_B T}).
\end{equation}
Unless stated otherwise, the free energy and the chemical potential will
be given in units of $k_B T$ and shifted by $\log(\lambda_B^2) -1$
in the following.

In practice, it is useful to expand functions of orientation $\vec{w}$
in spherical harmonics. Moments up to $l=10$ were taken into account,
{\rm i.e.}, 121 functions, a number which proved sufficient. The 
$z$-direction was discretized in steps of $l_0/5$. The mean field equations 
were solved iteratively, using the Legendre coefficients of the fields 
$h^{ind}_{h,t}(\vec{r},\vec{w})$ as iteration variables. The iteration
procedure combines a method proposed by Ng \cite{Ng} and simple mixing:
Let the vector $\vec{x}_n$
be the $n$th guess of the set of iteration variables, $\vec{f}_n$
the fields calculated from there, and $\vec{d}_n = \vec{f}_n - \vec{x}_n$
the remaining deviation. Following Ng, we define the matrix
$U_{ij}=(\vec{d}_{n}-\vec{d}_{n-i})\cdot (\vec{d}_{n} - \vec{d}_{n-j})$ 
and the vector $V_j = (\vec{d}_n-\vec{d}_{n-j}) \cdot \vec{d}_n$, where the
dimension of $U$ and $V$, $i_{max}=j_{max}$, is arbitrary (2 to 5 in this work).
We then invert $U$, determine the coefficients $A_i = U_{ij}^{-1} V_j$, and
calculate $\vec{x}_n^A = \vec{x}_n + \sum_i A_i (\vec{x}_{n-i} - \vec{x}_n)$ and
$\vec{f}_n^A = \vec{f}_n + \sum_i A_i (\vec{f}_{n-i} - \vec{f}_n)$.
In the iteration procedure suggested by Ng, the $(n+1)$th guess of $\vec{x}$
is given by $\vec{x}_{n+1} = \vec{f}_n^A$. 
Unfortunately, this method does not converge for the present problem. 
Good results were however obtained with the prescription
$\vec{x}_{n+1} = \vec{x}_n^A + \lambda (\vec{f}_n^A - \vec{x}_n^A)$,
with $\lambda$ ranging between $0.1$ and $0.2$. 
A relative accuracy of $10^{-8}$ was usually reached within less than 100 
iteration steps.
The iteratively obtained solutions for fixed surface coverage were usually 
unique, unless metastable states ({\em e.g.}, tilted states) existed. 
In that case, the solution with the lowest free energy (\ref{ffree})
was selected.

\bigskip

\section{Results.}

\bigskip

\subsection{Stiff Chains.}

Figure 3 shows a free energy curve in a system of relatively stiff 
chains ($u=2$, $v=13.7$). 
On increasing the molecular area, the free energy exhibits two minima
and then rises. As the area tends to infinity, not shown, it diverges 
negatively, following the ideal gas term $- \log(A/\lambda_B^2)$.
Hence the Maxwell enveloping function has a negative slope,
which guarantees the mechanical stability of the system: The
spreading pressure $
{\Pi}/{k_BT}=-{\partial F}/{\partial A}\: N^{-1}$
is always positive. The pressure in the gas phase is however very low, 
in the Maxwell construction the common tangent with a coexisting gas phase
is practically horizontal.
Figure 3 illustrates the situation where one has two distinct regions
of phase separation, first between a condensed phase (LS) and an expanded
phase (LE), and then between an expanded phase and the gas phase (G). The fact
that there is phase separation can be inferred from the Maxwell construction, 
which does not follow the free energy curve in those two regimes, and from the 
observation that the Gibbs free energy (\ref{gg}) is not a unique function 
of the chemical potential (Figure 3, inset).

If one decreases the chain stiffness $u$ or the chain anisotropy $v$, the 
coexisting expanded and condensed phases merge into one at a critical point 
(Figure 4a and b). This point is difficult to locate from just looking at 
the free energy curves, but can be identified {\em via} the
inspection of the Gibbs free energy as a function of the chemical potential.
On increasing $u$ or $v$, on the other hand, the free energy minimum
belonging to the condensed phase decreases relative to the other minimum.
A triple point is encountered, beyond which the liquid expanded state
is metastable, and the condensed phase coexists with the gas phase. A state 
with collective tilt emerges in the unstable surface coverage region. 
Tilt order thus occurs in a system of fixed grafted chains, but is replaced 
by phase separation when the chains are allowed to move.

So far, our results essentially confirm and complete the results
reported in reference \cite{me1}. Systems of relatively stiff chains 
qualitatively show the same behavior as found earlier in a somewhat 
different model. Hence we shall not discuss this regime in more
detail. The phase diagram in the plane of anisotropy vs. molecular area
at chain stiffness $u=2$ is shown in Figure 5.

\subsection{Flexible Chains.}

The transition between the condensed phase and the expanded phase is 
governed by the interplay of chain flexibility and chain anisotropy. 
In systems of more flexible chains, one recovers two phase coexistence 
if the higher conformational entropy is compensated by higher 
effective segment anisotropy. 

Free energy curves for the set of parameters $u=1$, $v=20.3$, are shown
in figure 6. As in figure 3, there are two successive first order
transitions between fluid phases, passing from the gas phase (G) {\em via}
a liquid expanded phase (LE) to a liquid condensed phase (L${}_2$).
Contrary to the case of stiffer chains, however, the coexisting condensed 
phase is tilted. Upon further compression of the monolayer, an additional 
continuous transition to an untilted state (LS) takes place.

The tilt order can be measured in terms of the in-plane alignment of segments,
$d_{\parallel} = \sqrt{\langle w_x \rangle^2 + \langle w_y \rangle^2}$.
The fact that $d_{\parallel}\ne 0$ implies that the symmetry in the
$xy$ plane is broken. Figure 7 demonstrates that the ``tilted state'' indeed 
displays this kind of azimuthal order. In systems of fixed grafted chains,
the tilted state is stable in a coverage interval, bounded by a continuous 
transition at high coverage (marked I in figs 6-8) and by a first order 
transition to the untilted state at low coverage (marked II). When the chains 
are given lateral mobility, this second transition disappears in the 
coexistence region of the L${}_2$ and the LE phase.

Further insight can be gained from the inspection of the nematic order 
in the system. The relevant quantity here is the traceless ordering 
matrix \cite{degennes} ${\bf S} = \langle 3 w_i w_j - \delta_{ij} \rangle /2$.
It has the eigenvalues $\{S, -(S-\eta)/2, -(S+\eta)/2 \}$, with
the nematic order parameter $S$ and the biaxiality $\eta$.
The nematic order $S$ is always nonzero, since the chains are always
aligned to some extent in the direction perpendicular to the surface. 
As the molecular area increases, it decreases monotonically in both
the tilted and the untilted phase, yet it stays higher in the
tilted phase. At the first order transition (II), $S$ jumps from 0.37 
in the tilted phase to 0.21 in the untilted phase. 
The value of $S$ in the tilted state is thus comparable to its value 
in the nematic phase of liquid crystals, right at the transition to the
isotropic phase ($S=0.43$ in the Maier-Saup\'e model \cite{degennes}).
The in-plane symmetry breaking is reflected by the behavior of the biaxiality, 
which is nonzero only in the tilted state.

From these results the nature of the tilting transitions in the system can
be inferred. The discontinuous low coverage transition (II) is associated with
ordering/disordering of single segments. It is thus essentially a
a nematic-isotropic transition, analogous to those found in liquid crystals. 
The continuous high coverage transition (I), on the other hand, results
from in-plane ordering/disordering of whole chains. The surface induces
an orientation direction, hence the transition is of XY type \cite{plischke}. 
The two types of transitions are illustrated in figure 9.

The structure of the monolayer shall be analyzed in some more detail. The 
density profiles in the three phases do not differ remarkably
from each other. Examples are shown in figure 10. The total segment density
is constant throughout the layer and independent of the surface area per
chain or the tilt order. It is also independent of the chain stiffness
and chain anisotropy, and only determined by the interaction parameter $e$
(not shown). Compression results in thickening of the monolayer.

The distribution of bending angles $\theta$, shown in figure 11, is more
interesting. It has by construction two maxima, one at the bending angle
$\theta = 0$ and one at $\cos \theta = 0.5$. The area under the second
maximum gives the concentration of conformational (gauche) defects in the 
chains. As demonstrated in the inset, the distribution for the outermost angle,
the angle between the last two segments, is always the same
up to the molecular areas which were considered. The main graph shows the 
deviations from this distribution
for the inner angles. In the expanded phase, chains have more defects in 
the middle than at the ends, {\em i.e.} they are more disordered there.
In the condensed phases, in contrast, the conformational order is highest in 
the middle. This result is in agreement with molecular dynamics simulations 
\cite{moller,karaborni} and other model calculations \cite{rieu}.

We close this section with the discussion of the phase diagram at chain 
stiffness $u=1$ (figure 12). At low chain anisotropy $v$, there is only one
single untilted liquid phase, which coexists with the gas phase. A tilted phase 
L${}_2$ emerges at a tricritical point, $v=20.1$, and
separates two untilted liquid regions, the expanded (LE) and the
condensed phase (LS). The transition between the L${}_2$ and LS state is 
continuous at lower values of the anisotropy $v$, and replaced by phase 
separation at the tricritical point $v=22.7$. We note that the tilting 
transition in monolayers of chains with fixed homogeneous grafting density 
remains continuous. The tilt order parameter vanishes continuously at a 
critical line, which is however hidden in the coexistence region if 
the chains are mobile. Beyond the triple point, where the liquid 
expanded phase becomes metastable ($v=20.5$), the region of stability of the 
tilted phase narrows down and finally disappears at $v=34$. 

\subsection{Phase behavior.}

We have seen that systems of stiff chains exhibit fluid fluid coexistence
of two untilted liquid phases, whereas in systems of flexible chains, the 
liquid expanded phase coexists with a tilted condensed phase. 
In an intermediate range of stiffness, one can find both. An example is
the phase diagram for chains of stiffness $u=1.5$, shown in figure 13.
Phase separation between an expanded phase and an untilted condensed 
phase sets in at the critical point $v=16.8$. A tilted phase
emerges at $v=17.1$ in the coexistence region between the expanded
and the untilted phase. The liquid expanded phase ceases to be stable at 
the triple point $v=17.4$. The L${}_2$ phase and the LS phase are separated 
by a narrow coexistence region; in systems of slightly more flexible chains,
$u \le 1.45$, the transition can also be continuous in a window of $v$ 
(see figure 14).

Figure 14 summarizes the phase behavior for chain stiffnesses ranging between
between $u=1$ and $u=2$. It shows a projection of the three dimensional
phase diagram in the $(A,u,v)$ volume into the $(u,v)$ plane. 
The shaded area designates the region where a tilted L${}_2$ state is stable.
The transition from this phase to the untilted LS phase is continuous
in the light shaded area, and first order in the dark shaded region
({\em i.e.}, the two phases phase separate). The light shaded
area is thus bounded by two lines of tricritical points. The
liquid expanded phase (LE) is stable in the hatched area, which
is bounded by a tricritical or critical line, and a triple line.
At very large chain stiffness, $u \sim 3.5 $, these two lines merge and 
disappear (not shown). Hence we recover
the rigid rod result reported in the literature: Monolayers of rigid
rods display neither stable tilted phases nor a liquid expanded phase.

We shall comment on this diagram in a few points. 

First, monolayers of chains with fixed anisotropy $v$, $v > 16.4$, display 
tilted phases only if the chains are sufficiently flexible. This
substantiates our claim, that tilt order is stabilized 
by chain flexibility. 

Second, the untilted condensed phase and the expanded
phase, both fluid, are not fundamentally different from each other.
The possibility of, {\em e.g.}, hexatic order is ignored within our 
approximations. Such ordering has however been reported in the liquid
condensed phase of lipid monolayers \cite{helm87}, and is presumably present
in our model too. If this is indeed the case, the coexistence
between the LE phase and the LS or a L${}_2$ phase is expected to
end in a multicritical point, and the transition to turn into a
continuous transition at lower values of $u$ or $v$.

Third, the role of the temperature has to be discussed.
Assuming that the segment density in the monolayer
does not change much in the interesting temperature 
regime, the temperature enters mainly {\em via} the chain stiffness $u$ and 
the chain anisotropy $v$. These parameters contain the Boltzmann factor 
$1/k_B T$, and may have a complicated temperature dependence in addition.
Let us neglect the latter and take $u, v \propto 1/T$ for simplicity.
Under this assumption, $(v \cdot u)^{-1/2}$ is proportional to the temperature,
and $(v/u)^{1/2}$ is temperature independent. The second quantity is interesting
in it's own right, since it can be related to the chain length $n$: 
In a continuum approximation, where chains are treated as space curves of 
length $L$, $L\propto n$ and stiffness $\eta$, $eta\propto u$ with 
orientational dependent interactions $V$, $V\propto v$, it can be shown that 
only two of these parameters 
are independent, {\em e.g.}, $(u/n)$ and $(v n)$ \cite{me2}. 
Hence varying the chain length
$n$ has the same effect as varying $(v/u)^{1/2}$. One can speculate 
that this remains qualitatively true for discrete chains.

The different possibilities for temperature dependent phase behavior can be 
read off from figure 15, which redraws the diagram of figure 14 
in the axis variables $(v/u)^{1/2}$ and $(v \cdot u)^{1/2}$. 
For example, the phase diagrams at $(v/u)^{1/2}=3$ and $(v/u)^{1/2}=4$ 
resemble figs 5 and 12, respectively. In the neighborhood of the fluid-fluid
coexistence region, increasing the ``chain length'' variable $(v/u)^{1/2}$
produces almost the same effect than decreasing the temperature. 
This fits to the experimental observation that the addition of two 
(CH${}_2$) groups to a system has a comparable effect to the reduction of
the temperature by $10-20 {}^0C$ \cite{stenhagen,bibo90}.
At $(v/u)^{1/2}=3.4$, the phase behavior of figure 13
is recovered, which is similar to the experimental phase diagram
sketched in figure 1. 

Note that mean field theories generally overestimate transition
temperatures. The effect is particularly strong in two dimensional
systems, where the fluctuations even prevent the possibility of true
long range tilt order (\cite{mw}), and second order tilting
transitions are replaced by Kosterlitz Thouless type transitions.
Hence the phase diagrams cannot be expected to be quantitatively
correct, and Figure 15 gives just a qualitative picture of the
phase behavior. This picture could be tested in simulations,
by systematic variations of chain length and chain stiffness.

\bigskip

\section{Conclusions.}

\bigskip

We have discussed the interplay of chain anisotropy and conformational entropy 
in simple model systems for Langmuir monolayers: Systems 
of short chains, which are confined to a planar surface at one end. 
The phase behavior as a function of chain stiffness and effective anisotropic
interaction was calculated in mean field approximation. 

We found that systems of chains with fixed grafting points, {\em i.e.}, 
fixed homogeneous grafting density, display tilt order in a density interval. 
It is bounded by a continuous transition to an untilted phase at high 
coverage, and by a discontinuous transition at low coverage. 
The high coverage transition involves ordering of whole chains and is of 
XY type, the low coverage transition is caused by ordering of segments and is 
reminiscent of the nematic/isotropic transition in liquid crystals. 
Note that beyond mean field theory, long wavelength fluctuations of the 
direction of tilt destroy the long range tilt order \cite{mw}. 
However, one can still expect quasi long range order, 
{\em i.e.}, correlation functions decay algebraically.

If the chains are free to move in the plane, tilt order is replaced by phase 
separation in systems of stiff chains. In systems of flexible 
chains, tilted phases remain stable to some extent. The conformational
entropy of the chains stabilizes tilt order. In fact, it favors phases
at lower surface coverage in general, which engenders both tilted phases
and an additional untilted phase, the liquid expanded phase.

As a function of the chain stiffness (or, as we have argued, the chain length), 
one can distinguish between four different regimes.
\begin{enumerate}
\item[(a)] Very stiff chains (rigid rod limit): 
Only one first order order transition is found, from the
highly diluted gas phase to the untilted liquid condensed phase. 
\item[(b)] Stiff chains (or short chains):
An additional untilted phase appears in a temperature interval.
One finds two successive fluid fluid transitions from the gas phase,
passing the liquid expanded phase, to the liquid condensed phase.
\item[(c)] Chains of intermediate stiffness:
Tilted phases can be stable. Depending on the temperature, the liquid expanded 
phase coexists with either a tilted or an untilted condensed phase.
\item[(d)]  Flexible chains (or long chains): 
The liquid expanded phase coexists with a tilted condensed phase.
Upon compression of the monolayer, the tilted phase turns into
an untilted phase {\em via} a continuous or first order transition.
\end{enumerate}

Hence a complex phenomenology is found already in this simple model, which 
incorporates only a few aspects of the hydrophobic tails in amphiphilic
molecules, and entirely disregards the structure of the head groups. 
The different phases in Langmuir monolayers at low surface coverage
are largely recovered.

We conclude that the essential features of the phase behavior of 
Langmuir monolayers can already be produced by the alkane tails of
the surfactant molecules alone.
Nevertheless, the head groups have an important influence on the phase diagram.
For example, it has been mentioned, that tilted phases can be 
stabilized by a mismatch between head group and tail segment size. This is 
most likely the dominant tilting mechanism in monolayers of single chain
amphiphiles, {\em e.g.}, fatty acids. Future investigations
will have to explore this possibility.

\bigskip

\section*{Acknowledgement.}

I have greatly benefitted from discussions with M. Schick, K. Binder, 
P. Nielaba, H. Lange, C. Stadler, and A. Halperin.
P. Nielaba and K. Binder are gratefully acknowledged for practical
advice and careful reading of the manuscript.

\newpage

\section*{Figure Captions}

\begin{description}

\item[Figure 1:] Phase diagram of Langmuir monolayers at low surface
  coverage (schematic). 
  The liquid-gas coexistence region is represented in a compressed way 
  relative to the liquid expanded-liquid condensed coexistence region. Whether 
  the latter ends in an upper critical, or turns into a second order transition 
  (indicated by the dashed line) in a multicritical point, has yet to
  be established (after Ref. \cite{knobler90}). 

\item[Figure 2:] Schematic picture of the model. Inset shows functional
  form of the bending potential. 

\item[Figure 3:] Free energy per particle vs molecular area 
  at chain stiffness $u=2$ and anisotropy $v=13.7$. Thin line indicates
  the Maxwell construction. Inset shows the Gibbs free energy per
  area $g/(k_B T)$ vs. the chemical potential. 

\item[Figure 4:] Free energy per particle vs molecular area 
  (a) for $u=2$ and different values of $v$; (b) for $v=13.7$ and
  different values of $u$. In (a) different offset values have been
  subtracted from the free energy. A state with tilt order emerges
  at high chain stiffness or high anisotropy (dashed line).

\item[Figure 5:] Phase diagram in the plane of anisotropy $v$ and molecular 
  area $A$ at chain stiffness $u=2$.

\item[Figure 6:] Free energy per particle vs molecular area at
  chain stiffness $u=1$ and anisotropy $v=20.3$. Two solutions of
  the mean field equations are shown, one corresponding to an untilted
  state (thick solid line) and one describing a tilted state (dashed line).
  Thin line indicates the Maxwell construction.

\item[Figure 7:] In-plane alignment of segments $d_{\parallel}$ vs
  molecular area at $u=1$, $v=20.3$. Solid line corresponds to the
  untilted state, dashed line to the tilted state. At fixed grafting
  density, the tilted state is stable in the coverage region between
  I and II.

\item[Figure 8:] Nematic order parameter $S$ and biaxiality $\eta$ vs
  molecular area at $u=1$, $v=20.3$. Solid line shows results for the
  untilted state, dashed line for the tilted state. Also indicated are
  the locations of the tilting transitions I and II at fixed grafting
  density, and of the coexistence regions between liquid phases in 
  systems of mobile chains.

\item[Figure 9:] Types of tilting transitions (see text for explanation).

\item[Figure 10:] Density profiles of the monolayer in the direction $z$ 
  perpendicular to the interface, at $u=1$, $v=20.3$ in different phases
  (different molecular areas $A$).
  Long and short dashed lines show the center of mass densities of
  tail and head segments $\widehat{\rho}_{h,t}(z)$, respectively; 
  solid line shows the total segment density $\rho(z)$.

\item[Figure 11:] Difference between the distribution of bending angles 
  $P(\cos \theta)$ in the middle ({\em i.e.}, between second and third
  tail segment) and at the end of the chains, 
  $\Delta P_{mid}(\cos \theta) = P_{mid}(\cos \theta) - P_{end}(\cos \theta)$.
  Results are shown for the parameters $u=1$, $v=20.3$ and different
  states (stable or unstable) at different molecular areas $A$. 
  Inset shows the distribution $P_{end}$ of the outermost angle, which was
  identical in all cases. 

\item[Figure 12:] Phase diagram in the plane of anisotropy $v$ and molecular 
  area $A$ at chain stiffness $u=1$. Inset shows a blow-up of the region
  where the liquid expanded phase is stable.

\item[Figure 13:] Phase diagram in the plane of anisotropy $v$ and molecular 
  area $A$ at chain stiffness $u=1.5$. 

\item[Figure 14:] Projection of the phase diagram in anisotropy $v$,
  chain stiffness $u$ and molecular area $A$ into the $(u,v)$ plane.
  Short dashed lines indicate multicritical lines, long dashed line 
  critical lines, and the solid lines are triple lines, where three phases 
  can coexist as indicated. Shaded areas are parameter regions where
  tilted phases can be stable. Coexistence of liquid phases is found in the 
  hatched area (expanded phase and one of the condensed phases) and in the 
  dark shaded area (tilted and untilted condensed phase). See text for
  further explanation.

\item[Figure 15:] Same as figure 14, with different axis variables.
  See text for explanation.

\end{description}

\newpage
\pagestyle{empty}

\figpage{1}{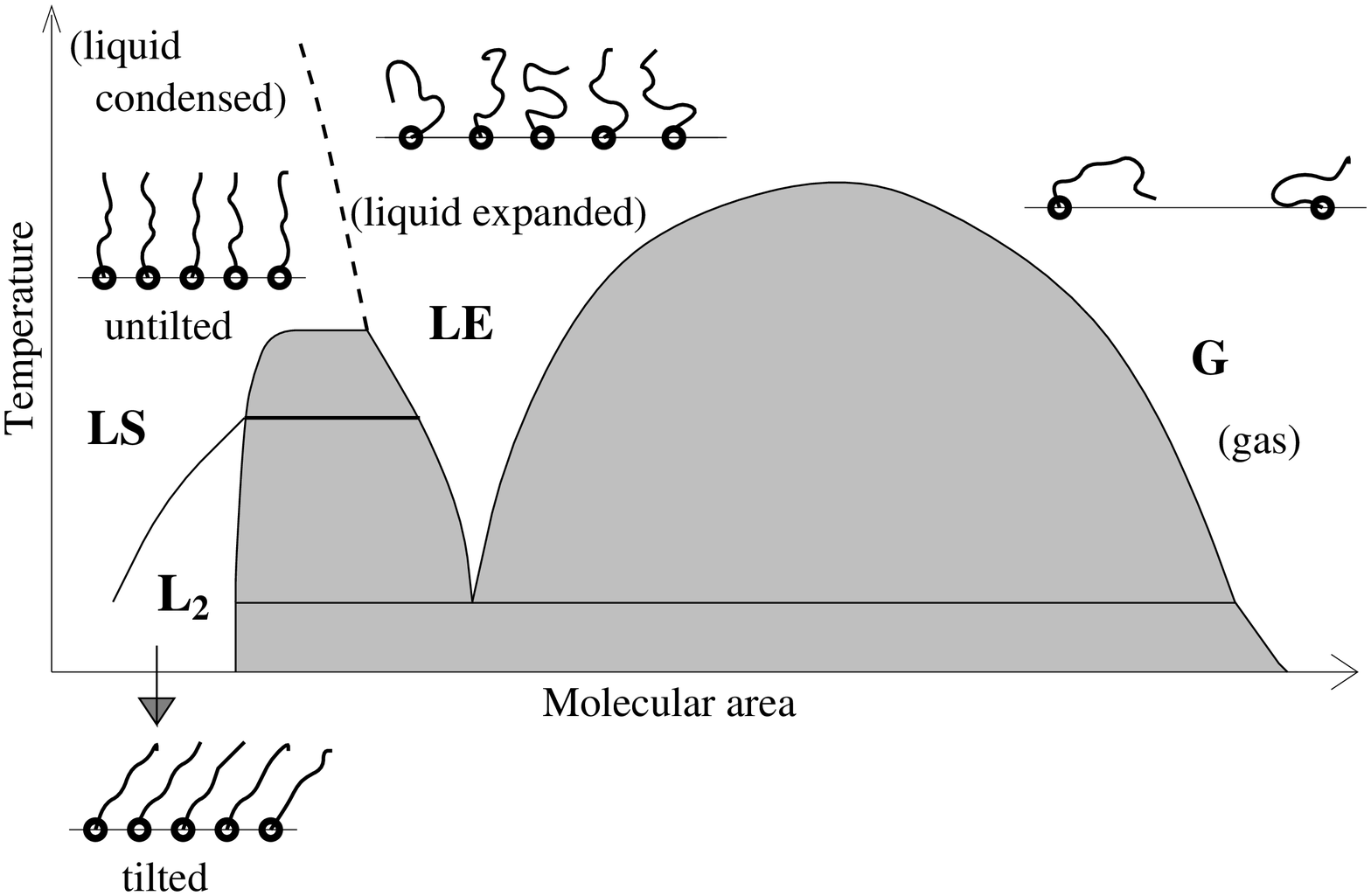}

\figpage{2}{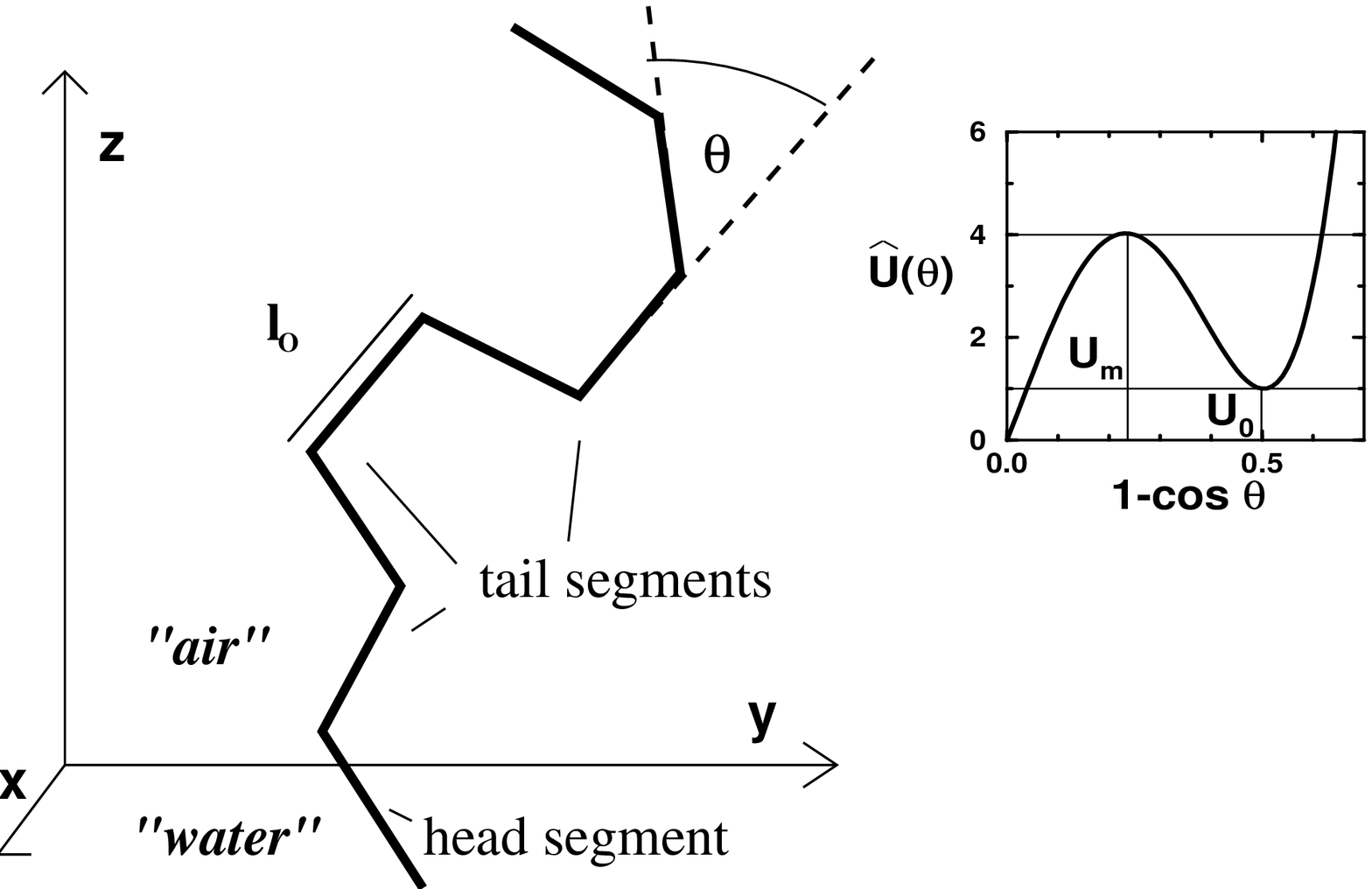}

\figpage{3}{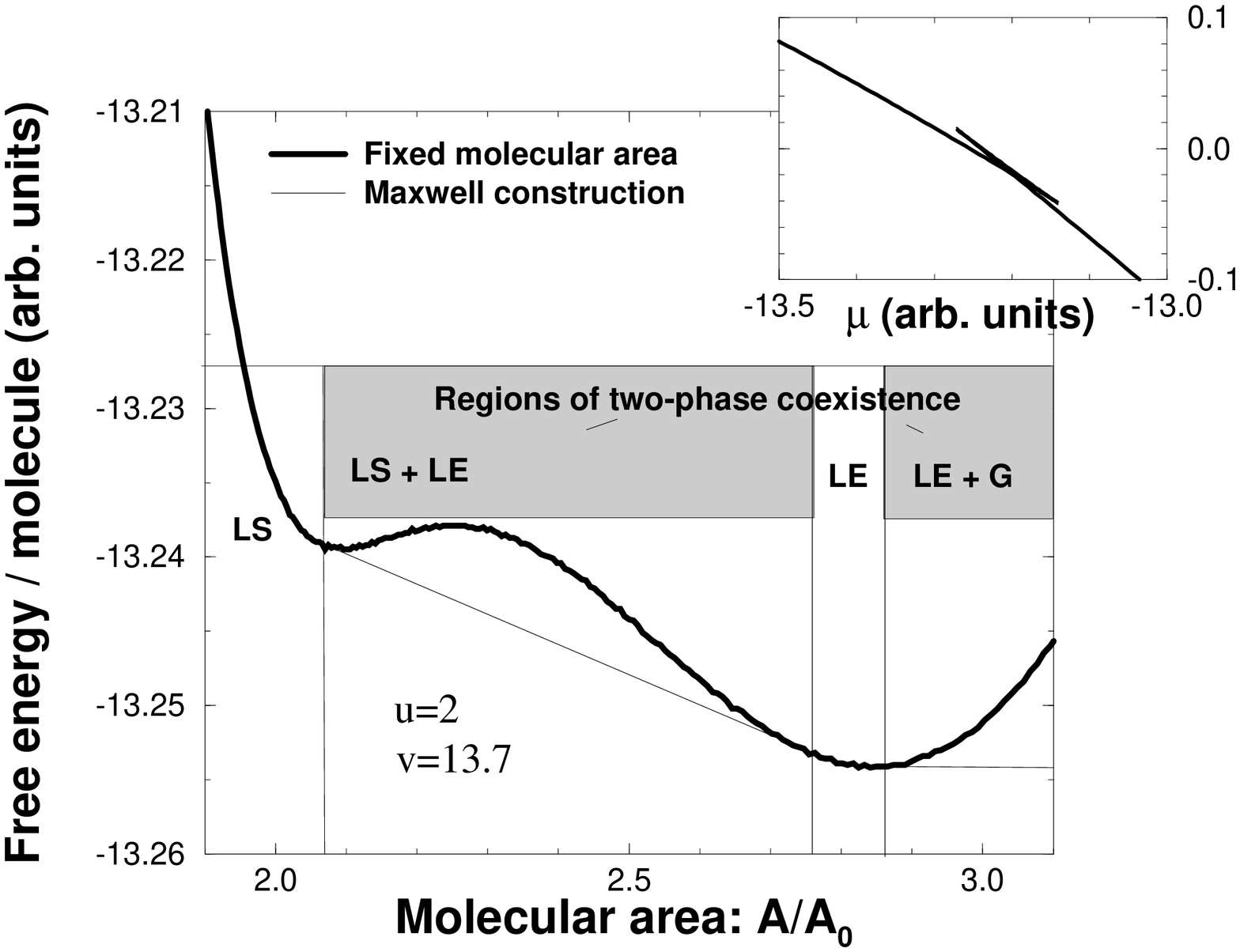}

\figpage{4a}{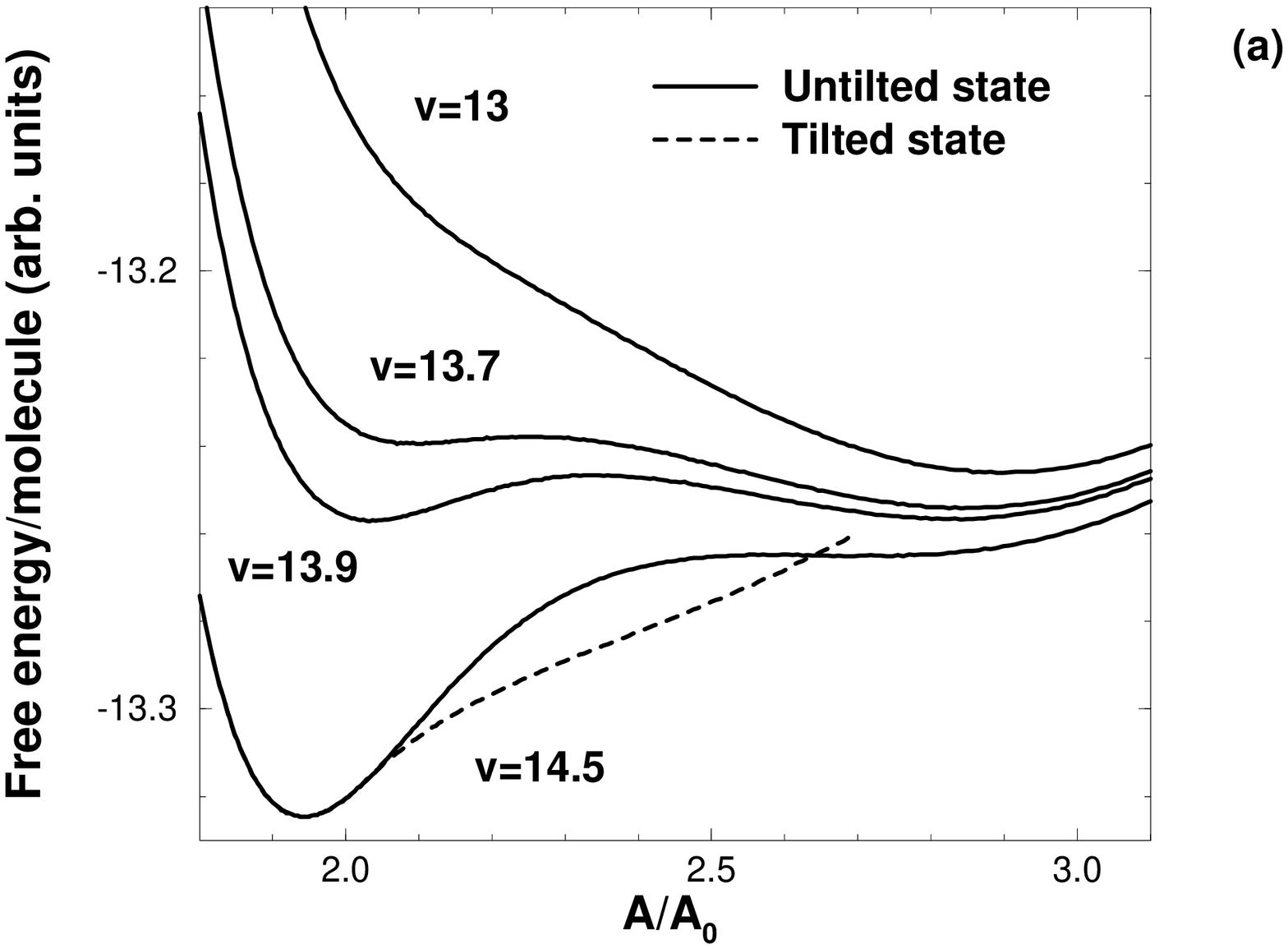}

\figpage{4b}{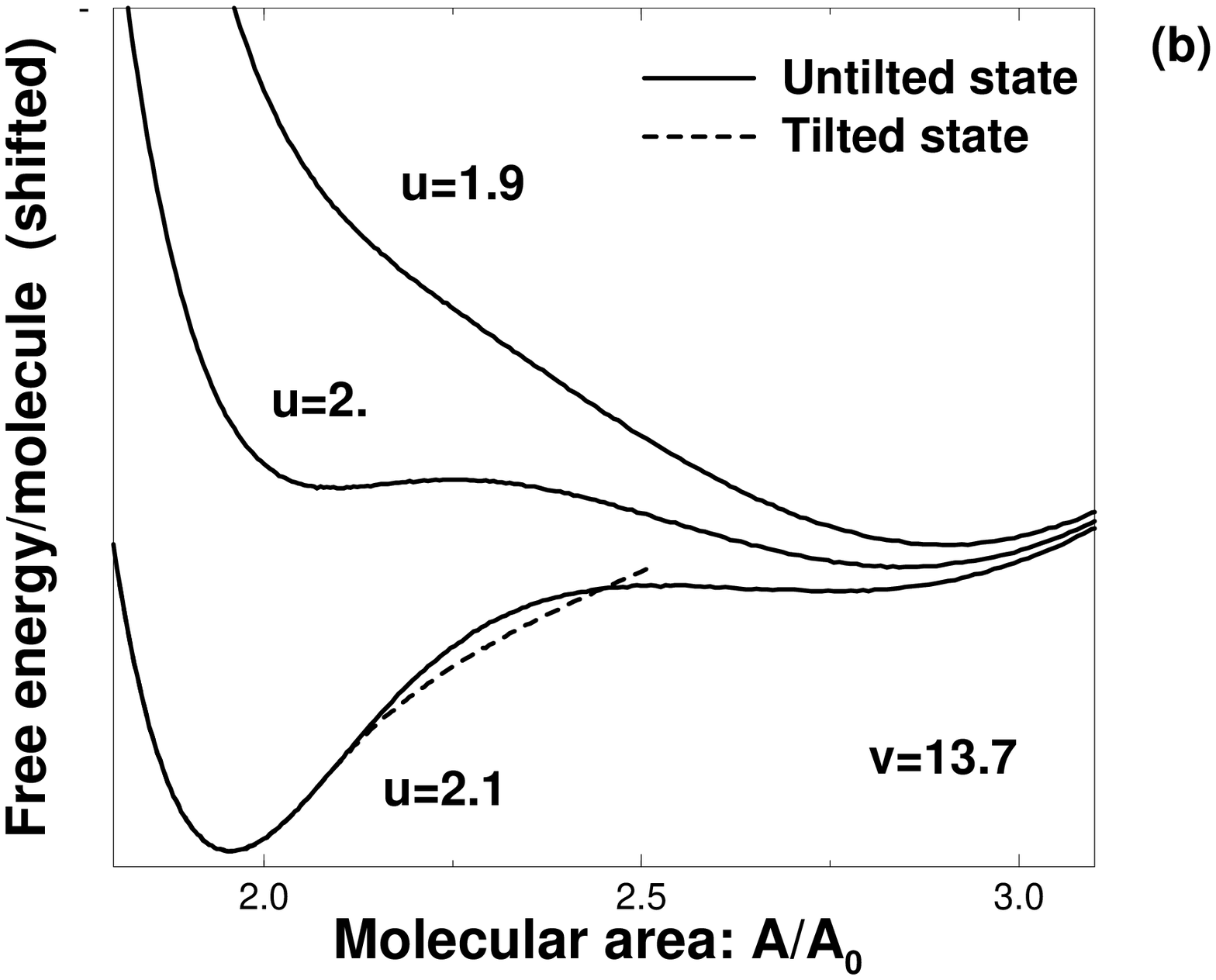}

\figpage{5}{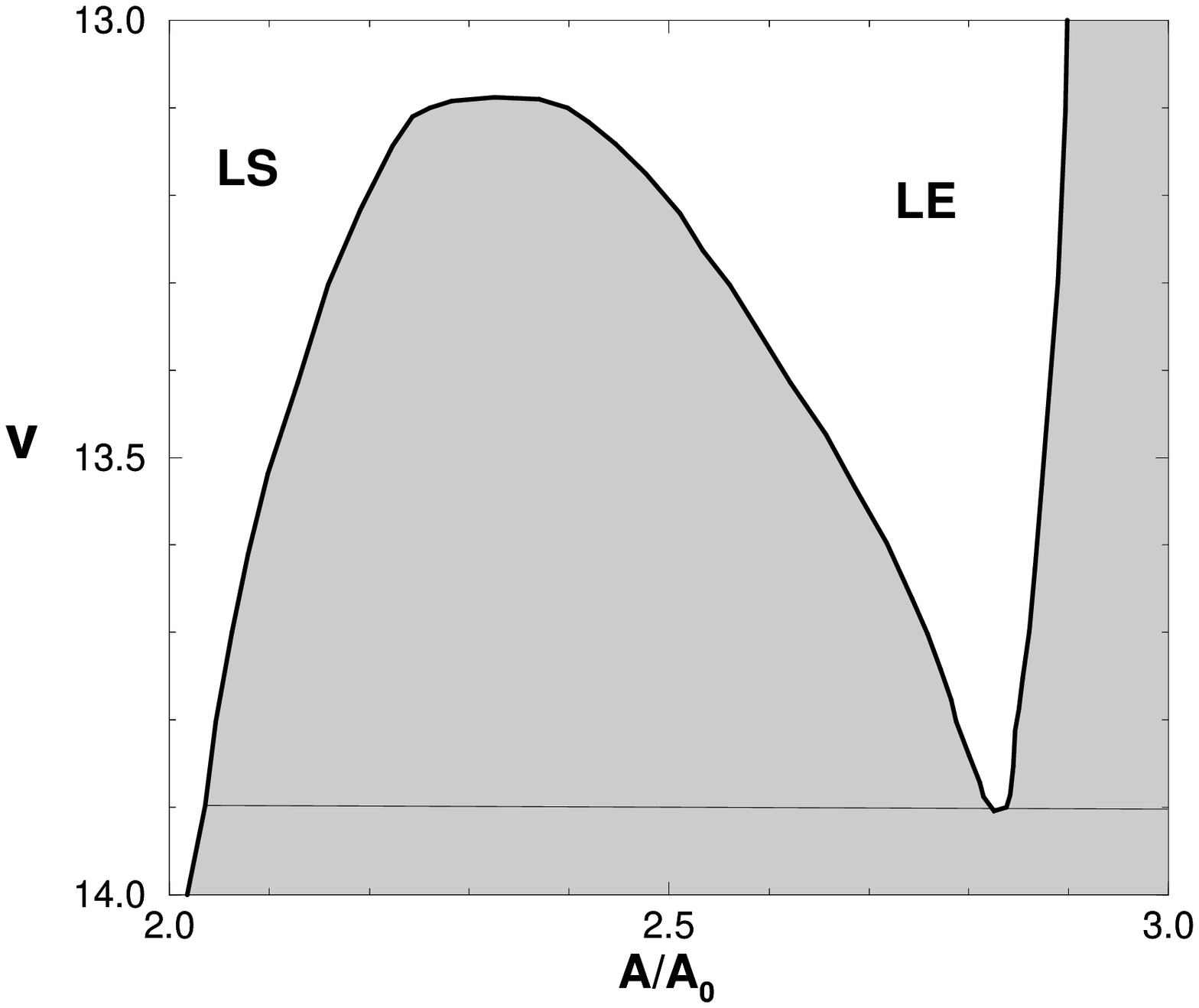}

\figpage{6}{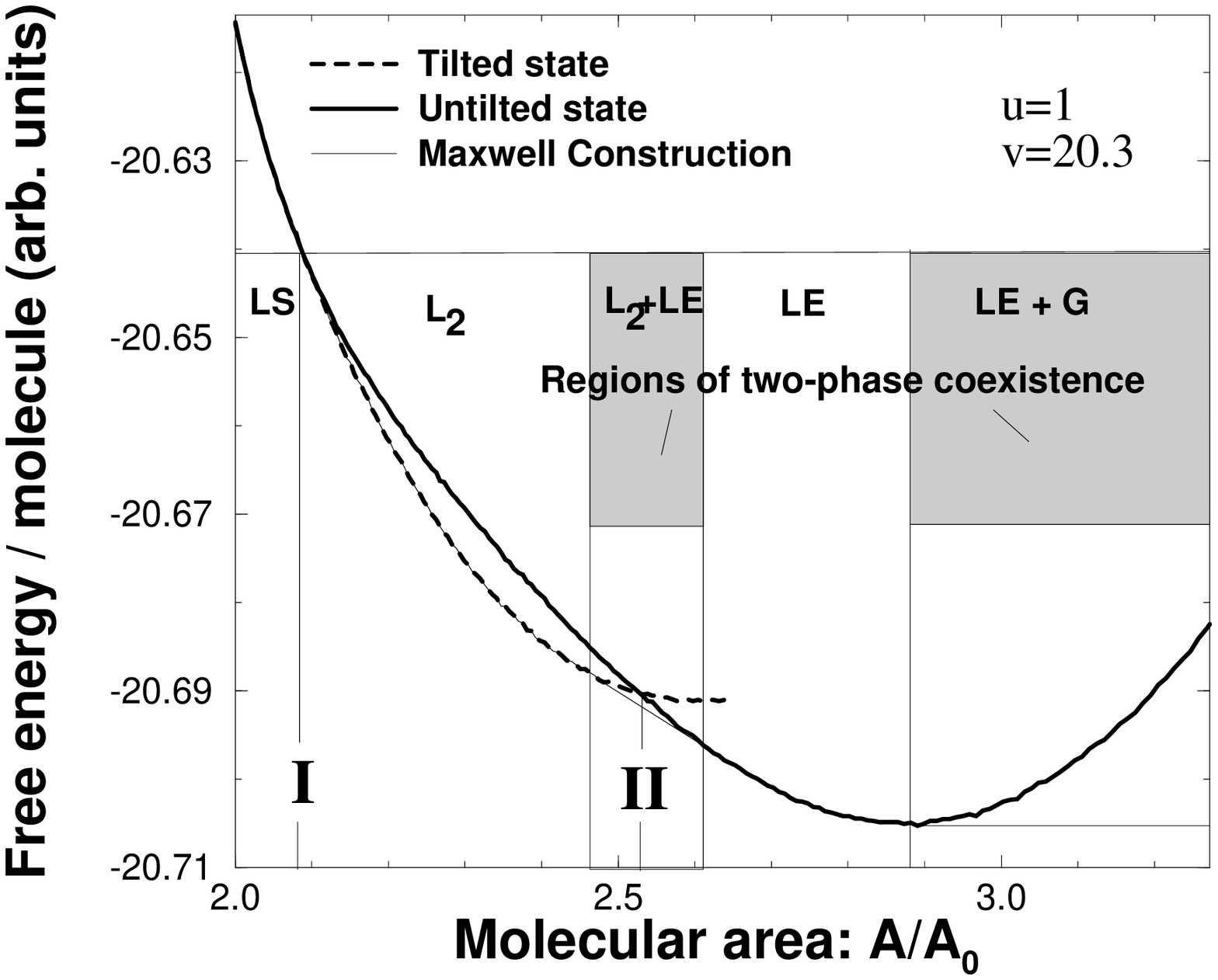}

\figpage{7}{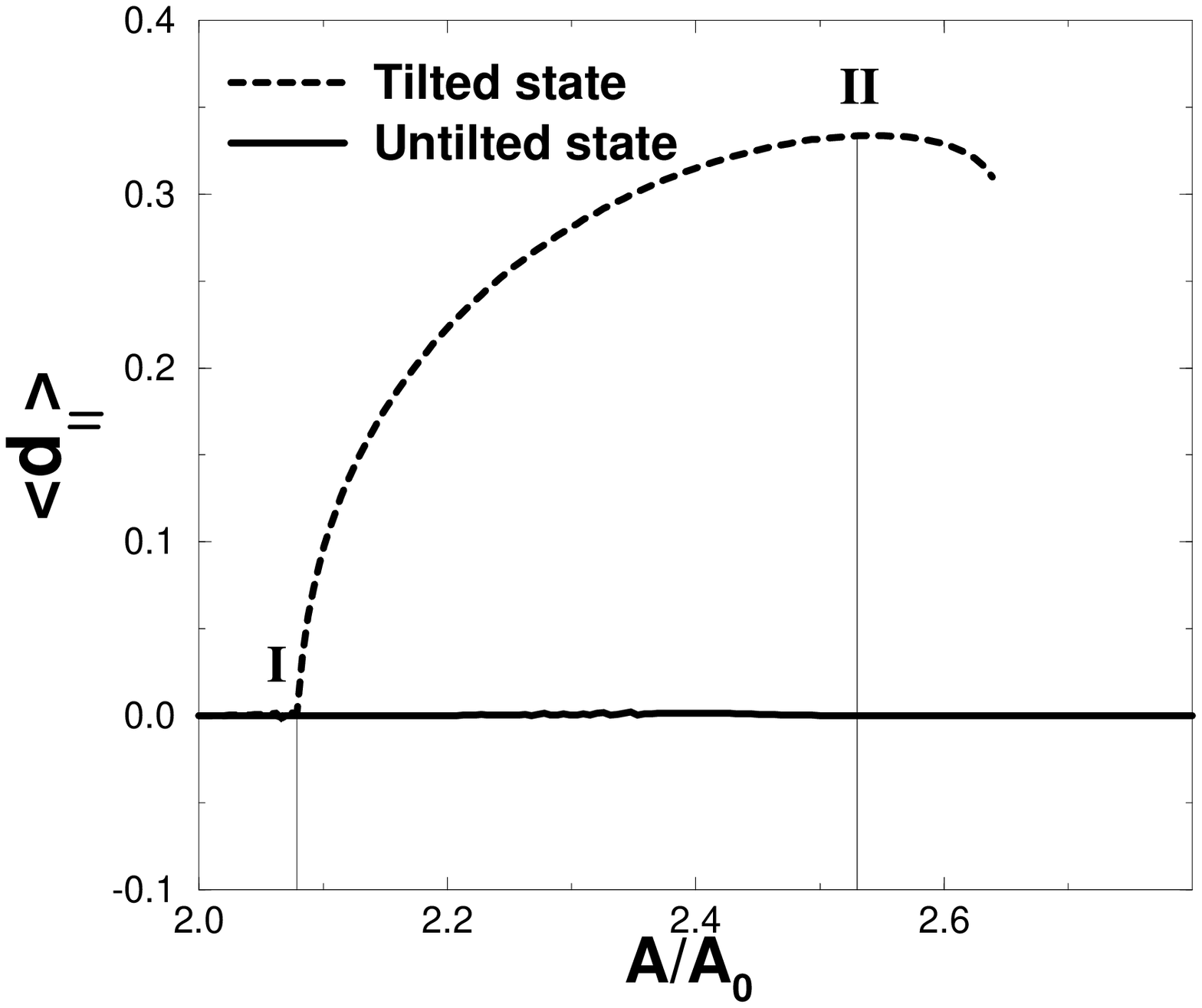}

\figpage{8}{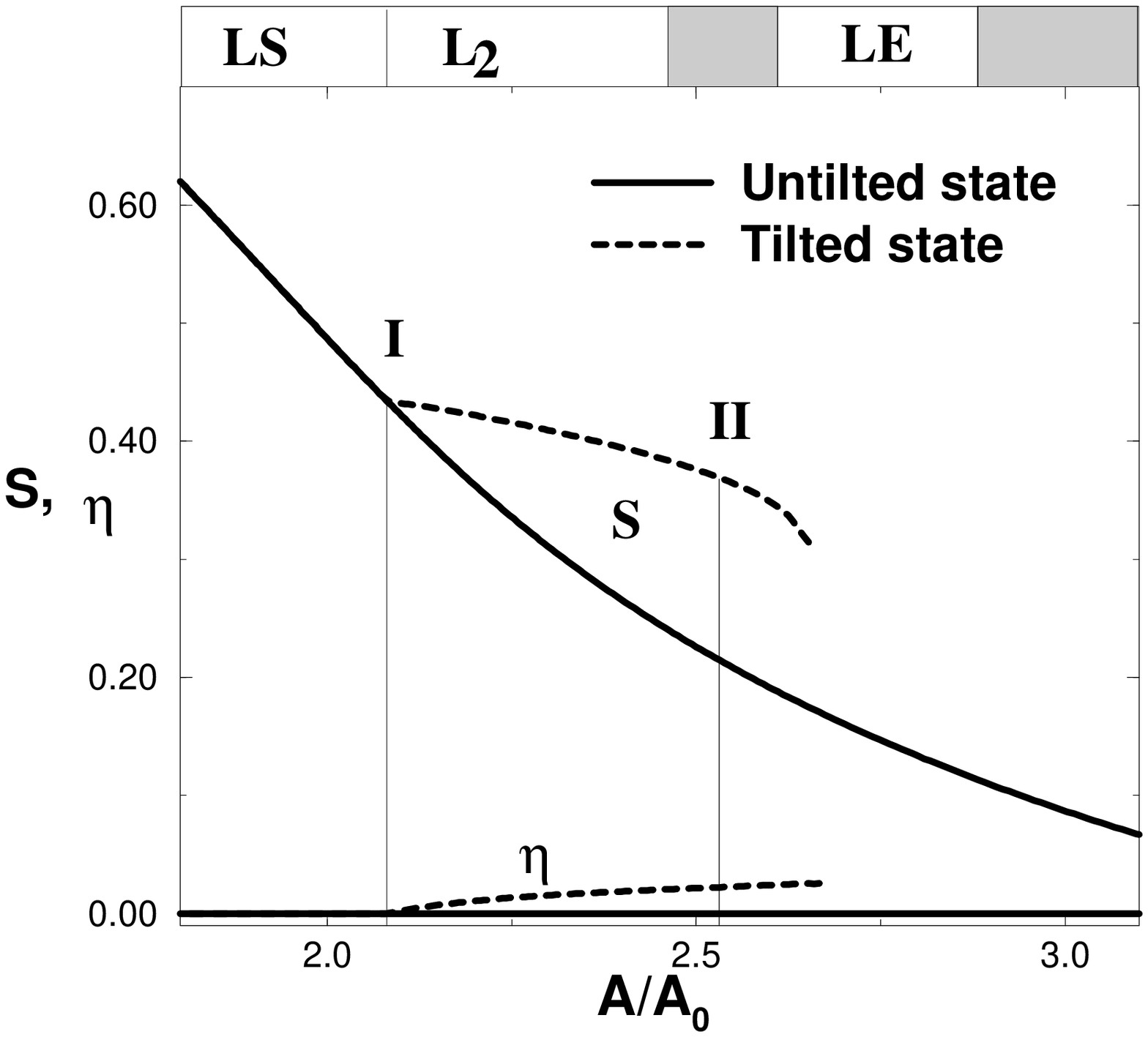}

\figpage{9}{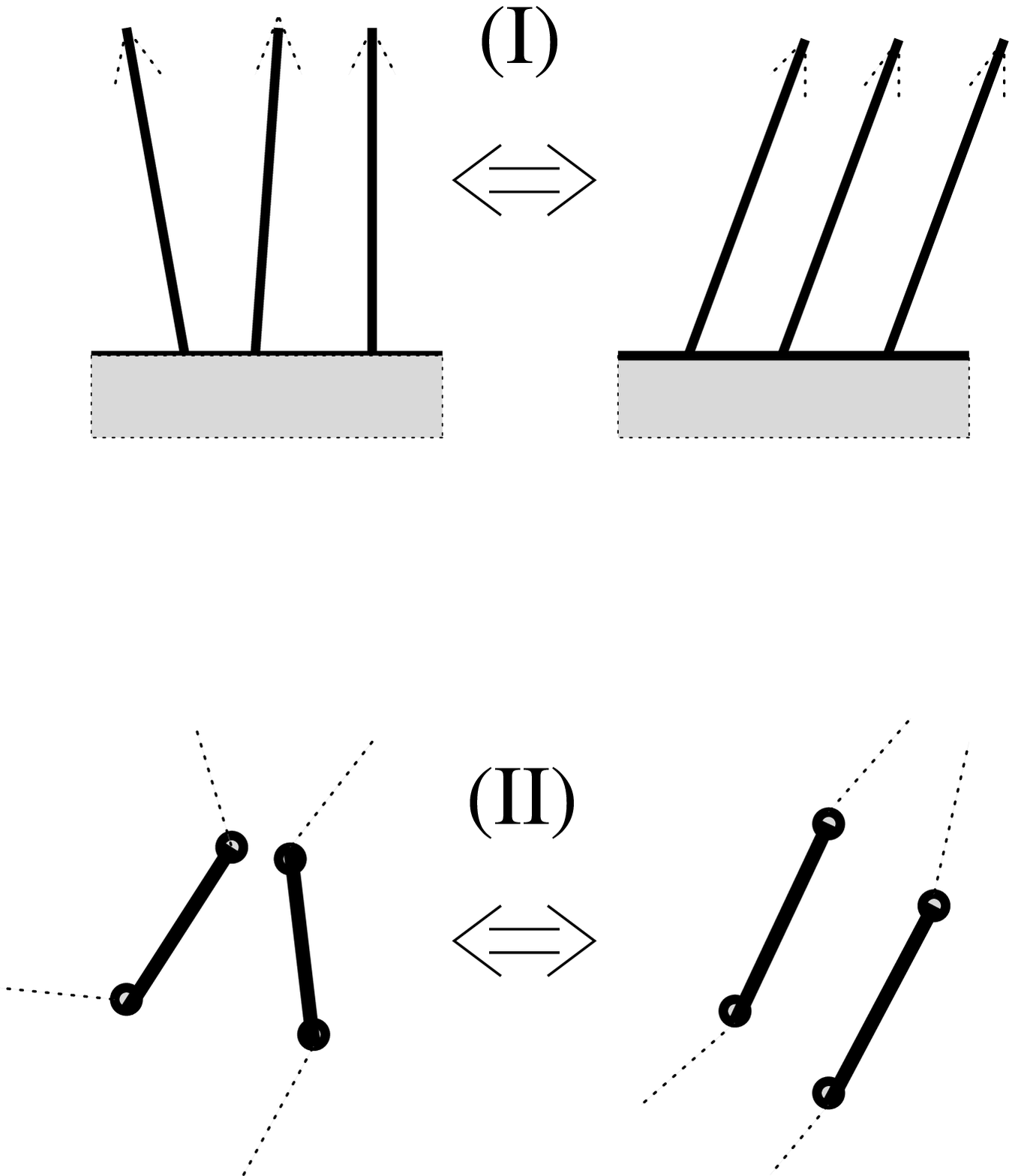}

\figpage{10}{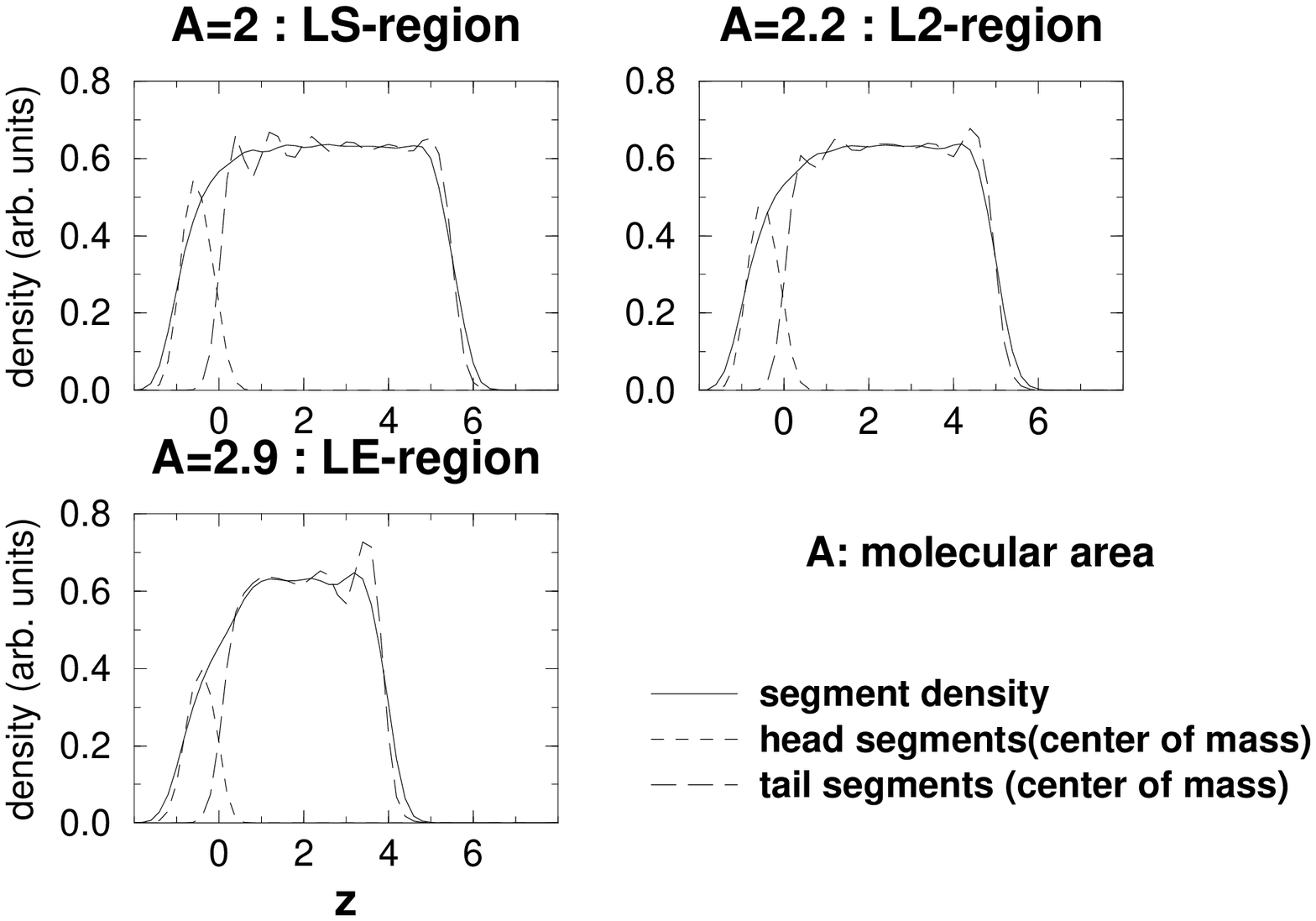}

\figpage{11}{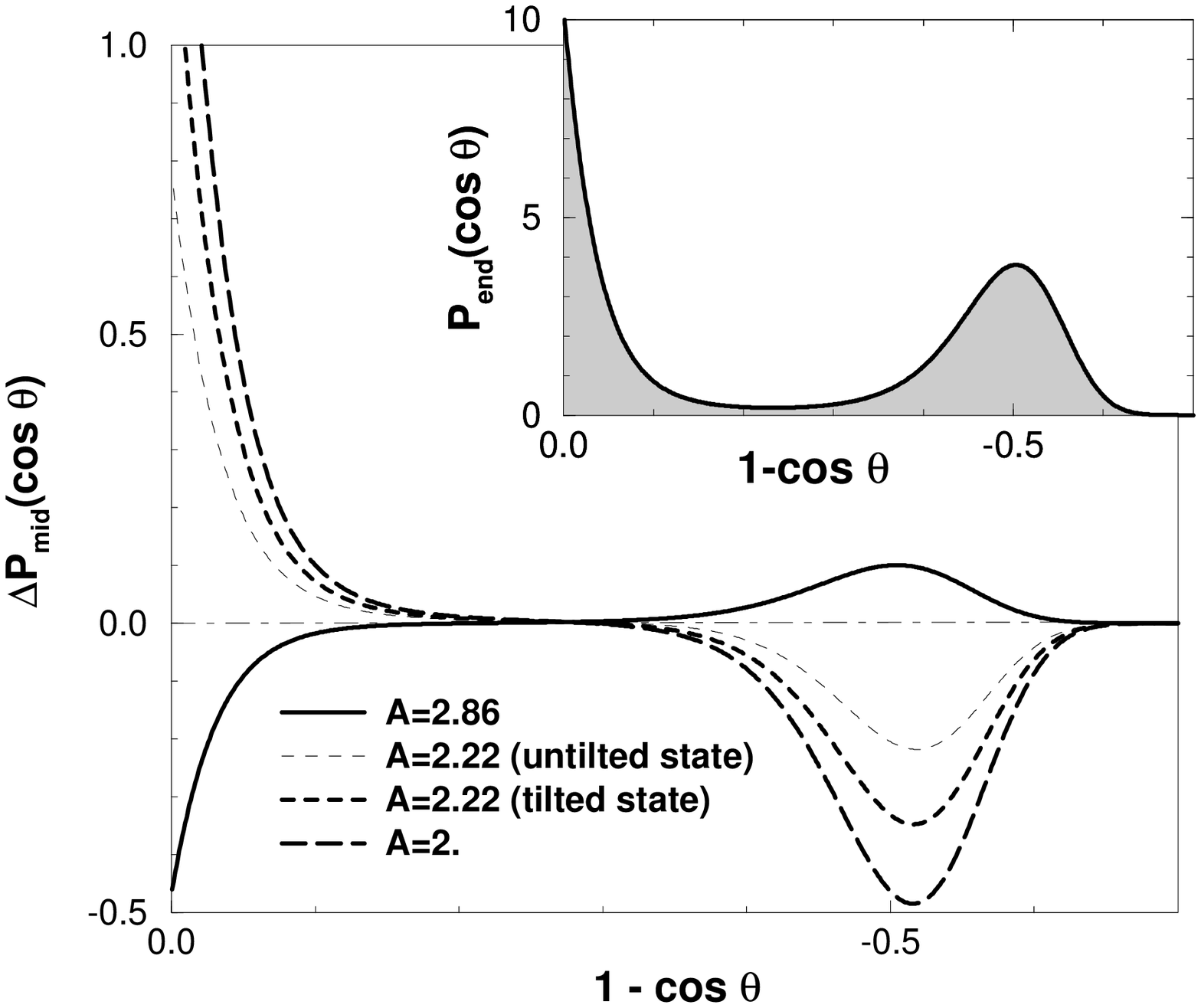}

\figpage{12}{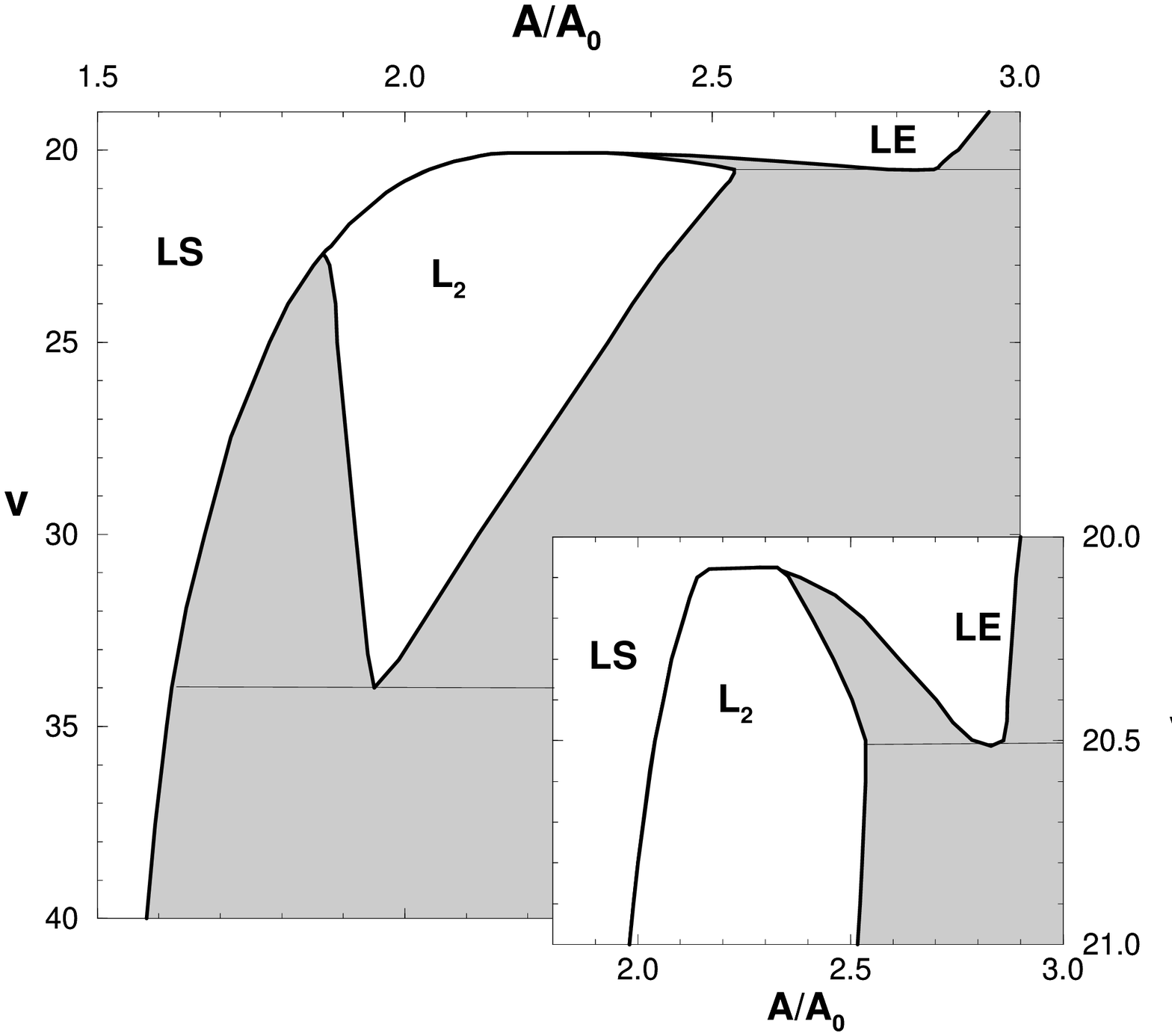}

\figpage{13}{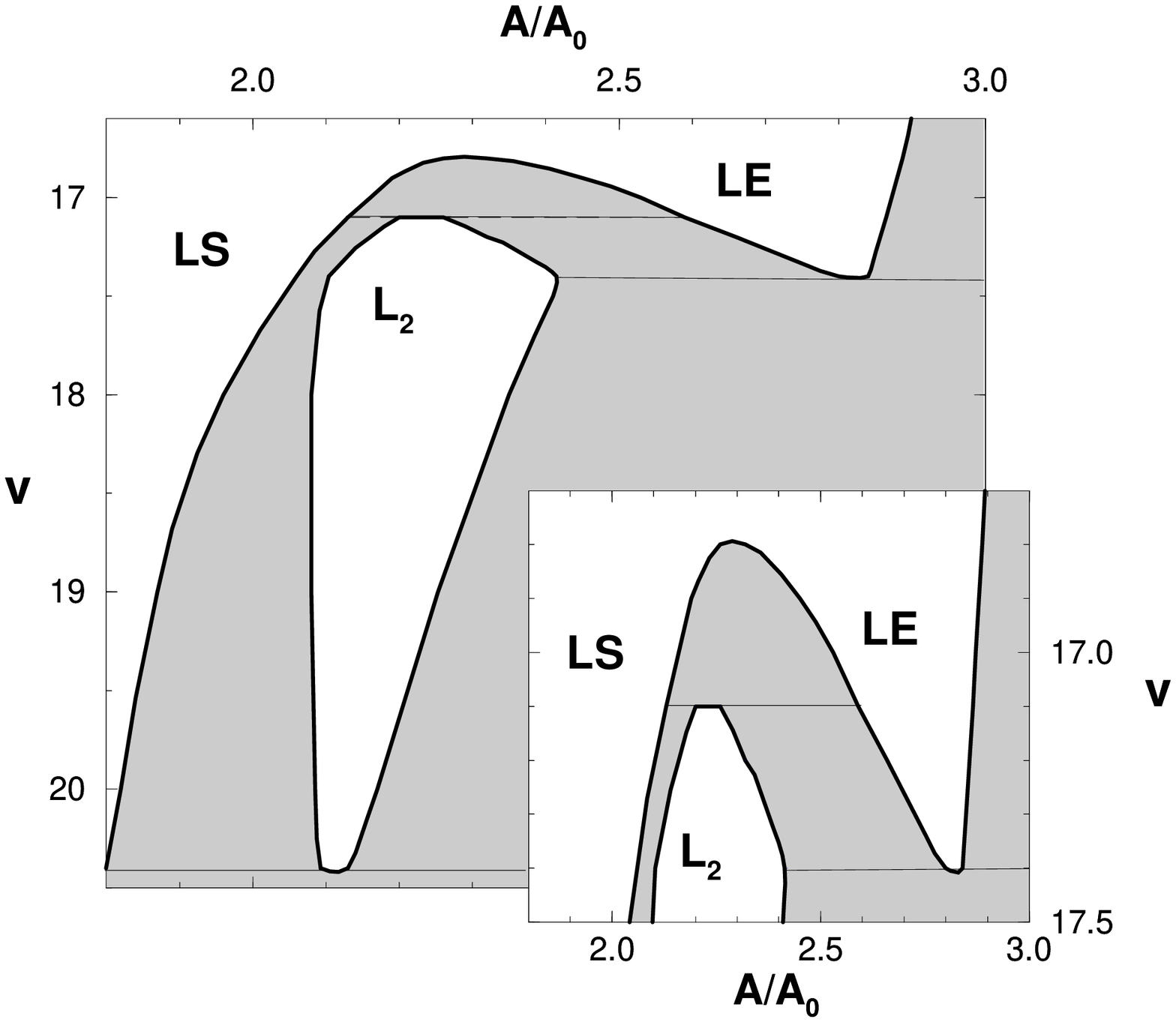}

\figpage{14}{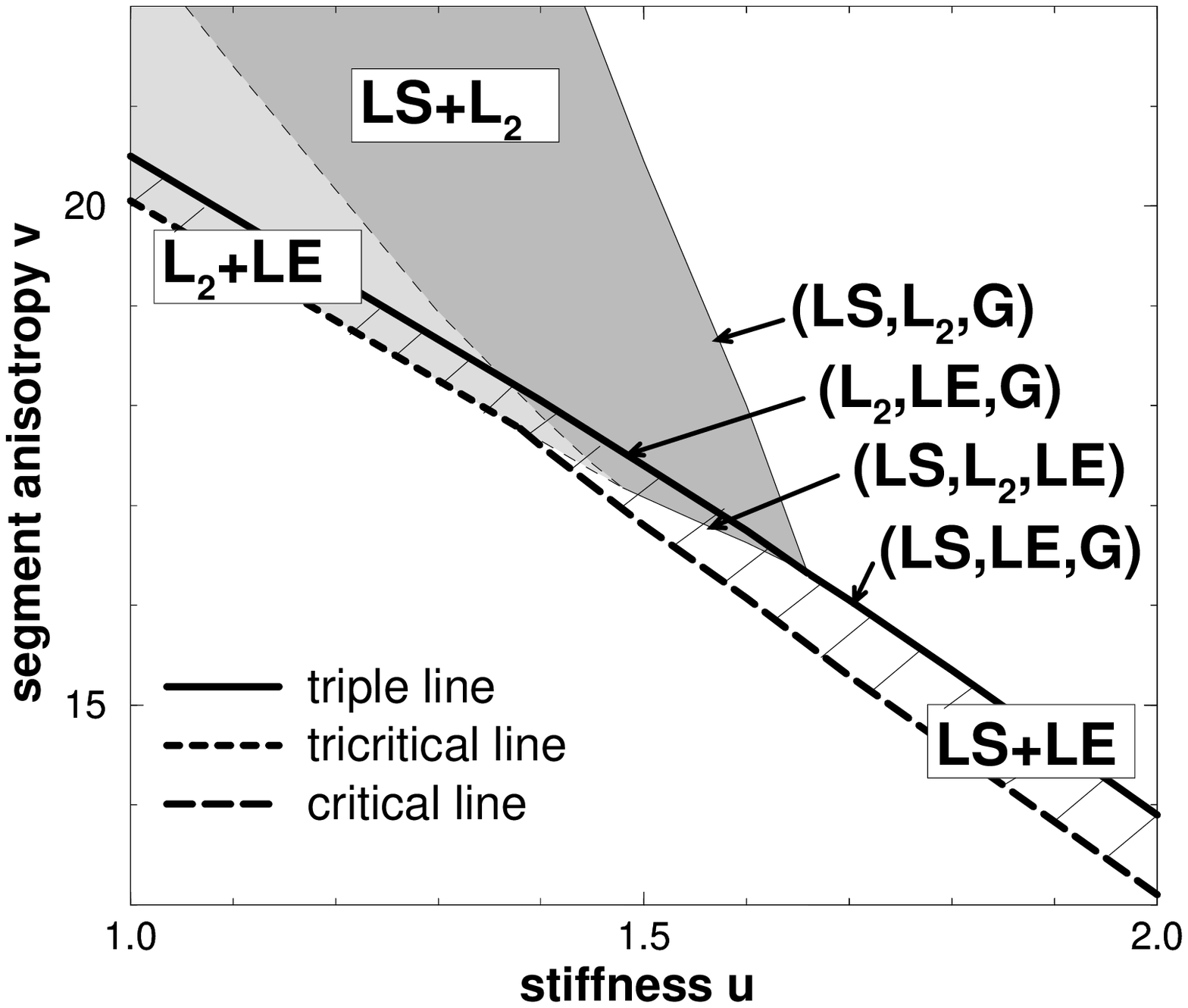}

\figpage{15}{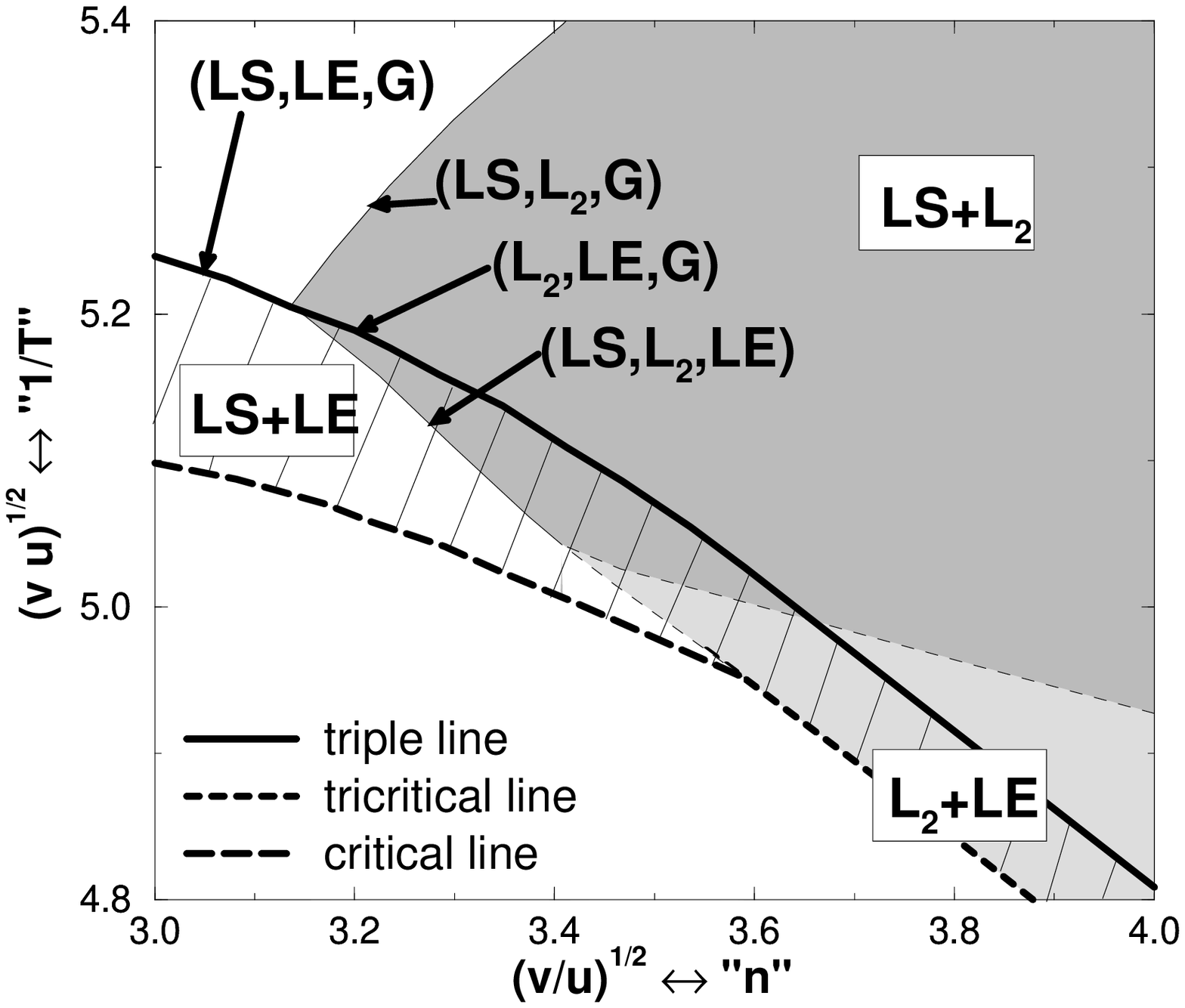}


\bigskip

\newpage

\begin{thebibliography}{99}

\bibitem{roberts} G.G. Roberts,
  Adv. Phys. {\bf 34}, 475 (1985).
\bibitem{gennis} R.B. Gennis: {\em Biomembranes}, Springer Verlag (1989).
\bibitem{knobler90} C.M. Knobler, Science {\bf 249}, 870 (1990).
\bibitem{mono-g} 
  H. M\"ohwald, Ann. Rev. Phys. Chem. {\bf 41}, 441 (1990);
  H.M. McConnell, ibid {\bf 42}, 171 (1991). 
\bibitem{helm87} 
  C.A. Helm, H. M\"ohwald, K. Kjaer, J. Als-Nielsen,
  Biophys. Journ. {\bf 52}, 381 (1987);
  K. Kjaer, J. Als-Nielsen, C.A. Helm, L.A. Laxhuber, 
  H. M\"ohwald, Phys. Rev. Lett. {\bf 58}, 2224 (1987).
\bibitem{bi} for a review on bilayers see, for example,
  M. Bloom, E. Evans, O.G. Mouritsen, Quart. Rev. Bioph. {\bf 24}, 293 (1991).
\bibitem{mono-p} 
  C.M. Knobler, R.C. Desai, Ann. Rev. Phys. Chem. {\bf 43}, 207 (1992).
\bibitem{kaganer} V.M. Kaganer, I.R. Peterson, M.C. Shih, M. Durbin, P. Dutta,
  J. Chem. Phys. {\bf 102}, 9412 (1995).
\bibitem{me1} F. Schmid, M. Schick, 
  J. Chem. Phys. {\bf 102}, 2080 (1995).
\bibitem{wang} Z. Wang, J. Physique France {\bf 51}, 1431 (1990).
\bibitem{scheringer} M. Scheringer, R. Hilfer, K. Binder, 
  J. Chem. Phys. {\bf 96}, 2269 (1991).
\bibitem{shin}
  S. Shin, N. Collazo, S.A. Rice, {\it J. Chem. Phys.} {\bf 98}, 3469 (1992);
\bibitem{li} M. Li, A.A. Acero, Z. Huang, S.A. Rice, 
  Nature {\bf 367}, 151 (1994).
\bibitem{dietrich}
  A. Dietrich, G. Brezesinski, H. M\"ohwald, B. Dobner, P. Nuhn,
  Il nuovo cimento {\bf 16D}, 1537 (1994).
  G. Brezesinski, C. B\"ohm, A. Dietrich, H. M\"ohwald,
  Physica B {\bf 198}, 146 (1994).
\bibitem{frank}
  F.M. Haas, R. Hilfer, K. Binder, J. Phys. Chem. (Sept. 1996, in press).
\bibitem{harald}
  H. Lange, diploma thesis.
\bibitem{Ahalperin}
  A. Halperin, S. Alexander, I. Schechter, 
  J. Chem. Phys. {\bf 91}, 1383 (1981).
\bibitem{karaborni2}
  S. Karaborni, G. Verbist,
  Europhys. Lett. {\bf 27}, 467 (1996).
\bibitem{li2}
  M. Li, S.A. Rice, 
  J. Chem. Phys. {\bf 104}, 6860 (1996).
\bibitem{rigby}
  D.J. Rigby, R.J. Roe, 
  J. Chem. Phys. {\bf 87}, 7285 (1987).
\bibitem{RIS}
  D.Y. Yoon, P.J. Flory,
  J. Chem. Phys. {\bf 61}, 5366 (1974).
\bibitem{scheutjens} J.M.H.M. Scheutjens, G.J. Fleer,
  J. Phys. Chem. {\bf 83}, 1619 (1979).
\bibitem{szleifer} I. Szleifer, M.A. Carignano,
  in {\em Advances in Chemical Physics}, Vol. XCIV, p. 165,
  I. Prigogine and S.A. Rice edts, Wiley (1996).
\bibitem{coy2}
  S.K. Nath, J.D. McCoy, J.P. Donley, J.G. Curro, 
  J. Chem. Phys. {\bf 103}, 1635 (1995).
\bibitem{me3} F. Schmid,
  J. Chem. Phys. {\bf 104}, 9191 (1996).
\bibitem{disks} E. Helfand, H.L. Frisch, J.L. Lebowitz, 
  J. Chem. Phys. {\bf 34}, 1037 (1960).
\bibitem{coy3} J.D. McCoy, S. Mateas, M. Zorly, J.G. Curro,
  J. Chem. Phys. {\bf 102}, 8635 (1995).
\bibitem{Ng} K.-C. Ng,
  J. Chem. Phys. {\bf 61}, 2680 (1974).
\bibitem{degennes} 
  P.G. DeGennes and J. Prost, 
  {\it The Physics of Liquid Crystals} Clarendon Press, Oxford, 2nd edn (1993).
\bibitem{plischke}
  M. Plischke, B. Bergersen,
  {\it Equilibrium Statistical Physics} 
  World Scientific, Singapore, 2nd edn (1994).
\bibitem{moller} M.A. Moller, D.J. Tildesley, K.S. Kim, N. Quirke,
  J. Chem. Phys. {\bf 94}, 8390 (1991).
\bibitem{karaborni} S. Karaborni, S. Toxvaerd,
  J. Chem. Phys. {\bf 96}, 5505 (1992); {\bf 97}, 5876 (1992);
  S. Karaborni, S. Toxvaerd, O. Olsen,
  J. Phys. Chem. {\bf 96}, 4965 (1992);
  S. Karaborni, Langmuir {\bf 1993}, 1334 (1993).
\bibitem{rieu}
  J.P. Rieu, M. Vallade,
  J. Chem. Phys. {\bf 104}, 7729 (1996).
\bibitem{me2}
  F. Schmid, M. M\"uller,
  Macromolecules {\bf 28}, 8639 (1995).
\bibitem{stenhagen} S. St\"allberg-Stenhagen, E. Stenhagen,
  Nature {\bf 156}, 239 (1945).
\bibitem{bibo90} A.M. Bibo, I.R. Peterson, Adv. Mat. {\bf 2}, 309 (1990).
\bibitem{mw} N.D. Mermin, H. Wagner, 
  Phys. Rev. Lett. {\bf 17}, 1133 (1966).

\end{thebibliography}
\end{document}